\documentclass[journal]{IEEEtran}
\usepackage{blindtext,lipsum,letltxmacro,xparse}
\usepackage[ruled,vlined]{algorithm2e}
\usepackage{comment}
\usepackage{hyperref,cite,footmisc}
\usepackage{pifont}
\usepackage{url,color,multirow}

\usepackage{amssymb}
\usepackage{pifont}
\usepackage{booktabs}
\usepackage{amsmath}

\DeclareMathOperator{\TSE}{TSE}
\DeclareMathOperator{\enc}{Encoder}
\DeclareMathOperator{\dec}{Decoder}
\DeclareMathOperator{\extmix}{MixBlock}
\DeclareMathOperator{\exttgt}{TgtBlock}
\DeclareMathOperator{\adapt}{Adapt}
\DeclareMathOperator{\enrlnet}{EnrlBlock}
\DeclareMathOperator{\ExtractBlockI}{ExtractBlock1}
\DeclareMathOperator{\ExtractBlockII}{ExtractBlock2}

\DeclareMathOperator{\CE}{CE}
\def\y{{\mathbf y}}
\def\x{{\mathbf x}}
\def\xs{{\mathbf{x}^s}}
\def\xS{{\mathbf{x}^S}}
\def\b{{\mathbf b}}
\def\Z{{\mathbf Z}}
\def\Zs{{\mathbf Z^s}}
\def\z{{\mathbf z}}
\def\zs{{\mathbf z^s}}
\def\Ms{{\mathbf M^s}}
\def\Y{{\mathbf Y}}

\def\xhat{\hat{\mathbf x}^{s}}
\def\xhatS{\hat{\mathbf x}^S}

\def\o{\mathbf{o}^s}
\def\oS{\mathbf{o}^S}
\def\a{\mathbf{a}^s}
\def\astar{\mathbf{a}^{\tilde{s}}}
\def\estar{\mathbf{e}^{\tilde{s}}}
\def\A{\mathbf{A}^s}

\def\es{\mathbf{e}^s}
\def\eo{\mathbf{e}^{\text{class}, s}}
\def\eoS{\mathbf{e}^{\text{class}, S}}
\def\W{\mathbf W}
\def\ea{\mathbf{e}^{\text{enrl}, s}}
\def\eaS{\mathbf{e}^{\text{enrl}, S}}

\usepackage[acronym,shortcuts]{glossaries}
\glsdisablehyper
\newacronym{SDR}{SDR}{signal-to-distortion ratio}
\newacronym{SDRi}{SDRi}{\gls{SDR} improvement}
\newacronym{SNR}{SNR}{signal-to-noise ratio}
\newacronym{SI-SDR}{SI-SDR}{scale-invariant \gls{SDR}}
\newacronym{NN}{NN}{neural network}
\newacronym{NNs}{NNs}{\gls{NN}s}
\newacronym{mAP}{mAP}{mean average precision}
\newacronym{GAN}{GAN}{generative adversarial network}
\newacronym{PIT}{PIT}{permutation invariant training}
\newacronym{IS}{IS}{inactive samples}
\newacronym{DNN}{DNN}{deep \gls{NN}}
\newacronym{AE}{AE}{acoustic event}
\newacronym{SE}{SE}{sound event}
\newacronym{AED}{AED}{acoustic event detection}
\newacronym{AEC}{AEC}{acoustic event classification}
\newacronym{AT}{AT}{audio tagging}
\newacronym{SED}{SED}{sound event detection}
\newacronym{SEC}{SEC}{sound event classification}

\newacronym{TSE}{TSE}{Target sound extraction}
\newacronym{USS}{USS}{universal sound separation}
\newacronym{ConvTasNet}{Conv-TasNet}{fully-convolutional time-domain audio separation network}
\newacronym{STFT}{STFT}{short-time Fourier transform}
\newacronym{iSTFT}{iSTFT}{inverse \gls{STFT}}
\newacronym{FiLM}{FiLM}{Feature-wise Linear Modulation}
\newacronym{ROC}{ROC}{receiver operating characteristic}
\newacronym{AUC}{AUC}{area under the curve}
\newacronym{PANN}{PANN}{pretrained audio \gls{NN}}
\newacronym{FSD}{FSD}{Freesound dataset}

\usepackage[dvipsnames]{xcolor}

\newcommand\noteNI[1]{\textcolor{black}{#1}}
\newcommand\noteNII[1]{\textcolor{black}{#1}}

\newcommand\noteI[1]{\textcolor{black}{#1}}
\newcommand\noteII[1]{\textcolor{black}{#1}}
\newcommand\noteIII[1]{\textcolor{black}{#1}}

\usepackage{ifpdf}
\usepackage{ulem}

\ifCLASSINFOpdf
  \usepackage[pdftex]{graphicx}
\else
  \usepackage[dvips]{graphicx}
\fi

\usepackage{array}

\ifCLASSOPTIONcompsoc
\usepackage[caption=false,font=normalsize,labelfont=sf,textfont=sf]{subfig}
\else
\usepackage[caption=false,font=footnotesize]{subfig}
\fi

\makeatletter
\def\ps@IEEEtitlepagestyle{%
  \def\@oddfoot{\mycopyrightnotice}%
  \def\@oddhead{\hbox{}\@IEEEheaderstyle\leftmark\hfil\thepage}\relax
  \def\@evenhead{\@IEEEheaderstyle\thepage\hfil\leftmark\hbox{}}\relax
  \def\@evenfoot{}%
}
\def\mycopyrightnotice{%
  \begin{minipage}{\textwidth}
  \centering \scriptsize
  Copyright~\copyright 2022 IEEE.  Personal use of this material is permitted.  Permission from IEEE must be obtained for all other uses, in any current or future media, including reprinting/republishing this material for advertising or promotional purposes, creating new collective works, for resale or redistribution to servers or lists, or reuse of any copyrighted component of this work in other works.
  \end{minipage}
}
\makeatother

\begin{document}
%
\title{SoundBeam: Target Sound Extraction Conditioned on Sound-Class Labels and Enrollment Clues for Increased Performance and Continuous Learning}
%
%
%

\author{Marc Delcroix~\IEEEmembership{Senior Member,~IEEE}, Jorge Bennasar Vázquez, Tsubasa Ochiai~\IEEEmembership{Member,~IEEE}, \\
Keisuke Kinoshita~\IEEEmembership{Senior Member,~IEEE}, Yasunori Ohishi~\IEEEmembership{Member,~IEEE}, Shoko Araki~\IEEEmembership{Fellow,~IEEE}
\thanks{The authors are with NTT Corporation, Japan. Jorge Bennasar Vázquez worked on this project during an internship at NTT.}
\thanks{Manuscript Submitted to IEEE/ACM Trans. Audio, Speech, and Language Processing on Feb. 10th, 2022; revised on Jul. 13, 2022 and Aug. 29, 2022; accepted on Oct. 20, 2022.}}

\maketitle

\begin{abstract}
In many situations, we would like to hear desired sound events (SEs) while being able to ignore interference. \noteII{Target sound extraction (TSE) tackles this problem by estimating the audio signal of the sounds of target SE classes in a mixture of sounds while suppressing all other sounds}. We can achieve this with a neural network that extracts the target SEs by conditioning it on clues representing the target SE classes. Two types of clues have been proposed, i.e., target \textit{SE class labels} and \noteI{\textit{enrollment audio samples} (or audio queries), which are pre-recorded audio samples of sounds from the target SE classes}. Systems based on SE class labels can directly optimize embedding vectors representing the SE classes, resulting in high extraction performance. However, extending these systems to extract new SE classes not encountered during training is not easy. Enrollment-based approaches extract SEs by finding sounds in the mixtures that share similar characteristics to the enrollment audio samples. These approaches do not explicitly rely on SE class definitions and can thus handle new SE classes. In this paper, we introduce a TSE framework, SoundBeam, that combines the advantages of both approaches. We also perform an extensive evaluation of the different TSE schemes using synthesized and real mixtures, which shows the potential of SoundBeam.
\end{abstract}


%
\IEEEpeerreviewmaketitle

\section{Introduction}
\IEEEPARstart{H}{\lowercase{uman}} beings can listen to a desired sound within a complex acoustic scene consisting of a mixture of various \glspl{SE}. This phenomenon is called the cocktail party effect or selective hearing~\cite{cherry_1953}. For example, it enables us to listen to an interlocutor in a noisy cafe, focus on a particular instrument in a song, or notice a siren on the road. One of the long-term goals of speech and audio processing research is to reproduce the selective hearing ability of humans computationally. 

\begin{figure}
    \centering
    \includegraphics[width=0.5\textwidth]{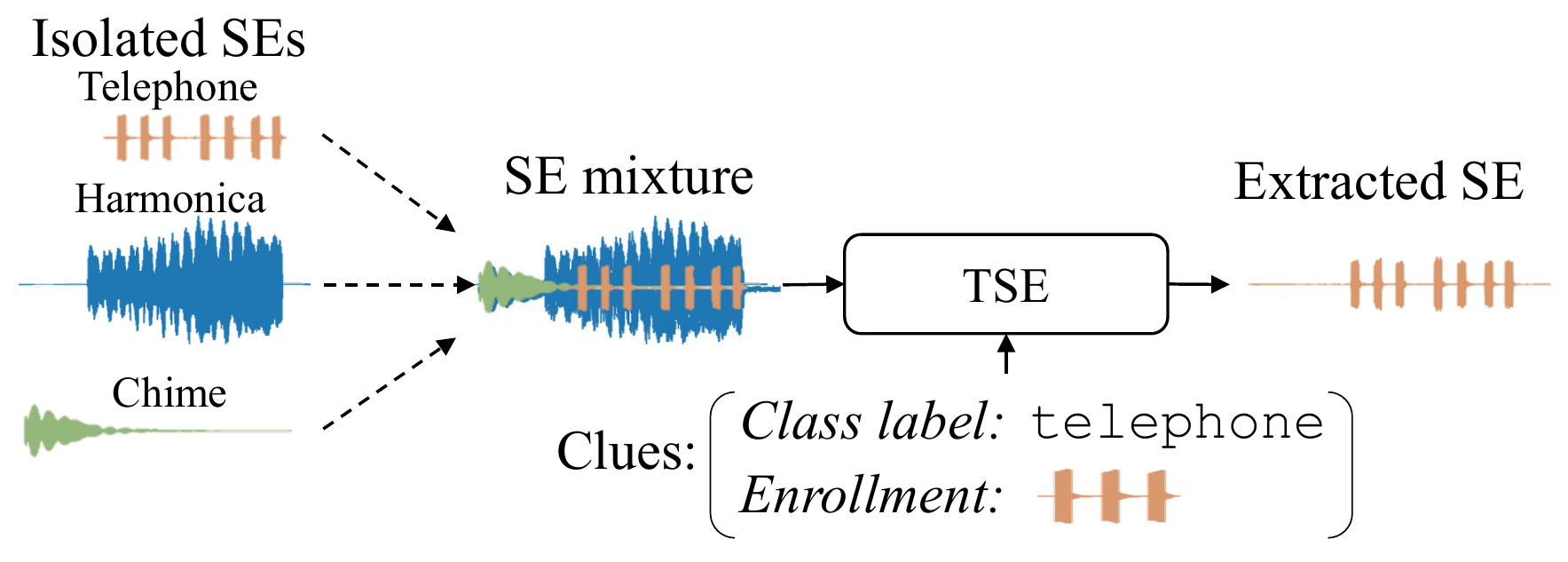}
    \caption{Schematic diagram of a \gls{TSE} system based on target \gls{SE} class labels or enrollment clues.}
    \label{fig:TSE}
\end{figure}

\noteII{\gls{TSE} is one approach toward achieving this goal}. 
\noteII{We define the \gls{TSE} problem as the extraction of one or multiple desired sounds from a mixture of various \glspl{SE}, given user-specified clues characterizing the target \gls{SE} classes. When multiple desired \gls{SE} classes are selected, we output a signal that consists of the sum of all the \glspl{SE} from these classes.}
\gls{TSE} exploits auxiliary clues such as class labels to inform the system about the target class~\cite{Ochiai2020}. \noteII{Figure~\ref{fig:TSE} is a schematic diagram of a \gls{TSE} system.}

In general, sound recordings are often polyphonic~\cite{cakir2015polyphonic} and contain different \glspl{SE} that overlap in time, which makes the problem particularly challenging. Solving the \gls{TSE} problem has many direct practical implications. For example, it would allow flexible personal hearables or \noteII{hearing aids} that could filter sounds depending on the situation to provide important information about our surroundings (e.g., a klaxon when we cross a street) while removing nuisances that disrupt our concentration (e.g., the same klaxon when we work at home with a window open). \noteNII{\gls{TSE} could also be used for sound post-production to emphasize or remove \cite{Ochiai2020} specific sounds in, e.g., a video recording.}

Research on the processing of \glspl{SE} has  focused mostly on \gls{SEC}, \gls{AT}, and \gls{SED} problems~\cite{mesaros2021sound,hershey2017cnn,kong2020panns,clavel2005events,cakir2015polyphonic, atrey2006audio,mesaros2017dcase}. \gls{SEC}/\gls{AT} assign \gls{SE} class labels to audio signals, while \gls{SED} also localizes in time the \glspl{SE} in the audio. \gls{SED} does not explicitly extract individual sound signals from the mixture, and thus it cannot solve the \gls{TSE} problem when \glspl{SE} overlap.

We can use source separation to separate a sound mixture into each source.   
For example, \gls{USS}~\cite{kavalerov2019universal} proposes to use a \gls{NN} to separate a mixture of arbitrary \glspl{SE}.
However, one issue with \gls{USS} is that it requires knowing or estimating the number (or maximum number) of sources in the mixture. This requirement may be challenging when dealing with a mixture of arbitrary \glspl{SE}, since the number of \glspl{SE} can vary greatly depending on the environment. Moreover, \gls{USS} provides access to the different sources  but does not identify them. Consequently, \gls{USS} is also unable to solve the \gls{TSE} problem. 

Recently, \gls{TSE} systems that can directly extract sounds from the target \gls{SE} classes have received increased interest~\cite{kong2020source,Ochiai2020,gfeller2020oneshot,delcroix2021few,okamoto2021environmental}. Such \gls{TSE} systems use a sound extraction \gls{NN} that accepts a sound mixture and an auxiliary target \gls{SE} embedding vector \footnote{\noteI{Some works call the ``target \gls{SE} embedding vector'' a ``conditioning vector'' \cite{kong2020source}, a ``conditioning embedding'' \cite{gfeller2020oneshot}, or a ``query embedding'' \cite{chen2022zero}.}} that represents the sound characteristics of the target \gls{SE}
. The sound extraction \gls{NN} extracts the sound signals in the mixture that match the characteristics of the target \gls{SE} embedding. The system thus implicitly performs ``separation'' and ``sound identification'' internally. 

There are several advantages to this approach. First, the sound extraction \gls{NN} extracts only the target \glspl{SE} and thus has a single output, which makes the \gls{NN} architecture independent of the number of sources in the mixture. Naturally, the computational complexity is also unaffected by the number of sources. Finally, the ``separation'' and ``identification'' processes are performed simultaneously, allowing us to exploit the information on the sound characteristics to help the internal ``separation'' process and optimize the \gls{TSE} system globally. 

A key component of \gls{TSE} systems are the target \gls{SE} embeddings. 
They can be derived from \textit{class labels} or \textit{enrollment audio samples}\footnote{\noteI{Some works use the term ``audio query''\cite{LeeCL19,chen2022zero} or conditioning audio \cite{gfeller2020oneshot} instead of ``enrollment audio samples.''}}.
With the first approach, class labels represented as 1-hot vectors are mapped to an embedding space representing the different \gls{SE} classes using an embedding layer, which is learned jointly with the sound extraction \gls{NN}~\cite{slizovskaia2019end,Ochiai2020,kong2020source}. Such class label-based \gls{TSE} systems can directly learn the target \gls{SE} embeddings, which can thus optimally represent the characteristics of the \gls{SE} classes for the \gls{TSE} task.
However, class label-based approaches assume a predetermined set of target \gls{SE} classes, and thus they cannot perform \gls{TSE} of \textit{new \gls{SE} classes} that were not encountered during training. 

With the second approach, the target \gls{SE} embeddings are derived from an enrollment audio sample, which is a short recording of an isolated sound similar to the target \gls{SE}~\cite{zmolikova2019speakerbeam,seetharaman2019class,LeeCL19,gfeller2020oneshot,delcroix2021few}.
Enrollment-based \gls{TSE} systems learn to extract sounds that share similar characteristics to the enrollment sample without explicitly relying on \gls{SE} class labels. They can thus perform \gls{TSE} of new \gls{SE} classes. However, unlike class label-based approaches, the embeddings are derived from the enrollment samples. Therefore, they are not directly optimized for the \gls{SE} classes, which may lead to sub-optimal performance.

In this paper, we provide a unified description of the class label and enrollment-based \gls{TSE} approaches and introduce \textit{SoundBeam}, which is a \gls{TSE} model that combines both approaches. \noteNI{SoundBeam generalizes the target \textit{speech} extraction framework called SpeakerBeam~\cite{zmolikova2019speakerbeam} to arbitrary \textit{sounds}. ``Beam'' refers to focusing on a particular sound in a mixture but does not imply using a microphone array or a beamformer~\cite{delcroix2018single}. Indeed, this work focuses on single-channel processing.}

SoundBeam learns an embedding space common to the class label- and enrollment-based target \gls{SE} embeddings by sharing the extraction \gls{NN}.
Consequently, we can use optimal \gls{SE} embeddings for known \gls{SE} classes with the class label embeddings and enrollment embeddings for new \gls{SE} classes. Furthermore, we show that SoundBeam has potential for \textit{continuous learning}, since it can learn class label embeddings for new classes with few-shot adaptation.

We presented the initial ideas of SoundBeam in our previous work~\cite{delcroix2021few}. This paper provides a more detailed explanation of the approach and an extensive evaluation of the \gls{TSE} frameworks, covering the following four practical aspects. 

\textbf{Extraction of new SE classes with few-shot adaptation}: 
    The number of potential \gls{SE} classes is huge, if not infinite, and therefore it may not be realistic to develop a system that can handle all possible \gls{SE} classes. Moreover, the target \gls{SE} classes may depend on the application scenarios. Therefore, we believe \gls{TSE} systems should be able to learn new \gls{SE} classes after their deployment, i.e., continuous learning. We propose a few-shot adaptation approach, which allows SoundBeam to learn optimal \gls{SE} embeddings from new \gls{SE} classes using a few enrollment samples. We introduced this adaption scheme earlier \cite{delcroix2021few}. Here, we provide a deeper experimental evaluation, including a class-wise analysis of the effect of adaptation and experiments on the influence of the number of enrollment samples on performance.
    
    \noteII{\textbf{Handling of inactive target \gls{SE} classes}}: 
    For many practical applications, a \gls{TSE} system should output a silent or zero signal if the target \gls{SE} is \textit{inactive} in the mixture. However, most prior works have not rigorously evaluated this scenario. We discuss how to learn to handle inactive \gls{SE} classes and perform a thorough experimental analysis with the different \gls{TSE} frameworks.
    
    \textbf{Simultaneous extraction of multiple \glspl{SE}:}
    In some applications, there may be several target \gls{SE} classes. The target would thus consist of a mixture of target \gls{SE} sounds. We show that we can extract several \gls{SE} sounds simultaneously simply by adding the \gls{SE} embeddings of the different target classes. We first introduced this idea for class label-based \gls{TSE}~\cite{Ochiai2020}. Here, we develop it for the enrollment and SoundBeam models.
    
    \textbf{Challenges faced with real recordings:}
    Finally, we explore the challenges of processing real mixtures, which include evaluation without access to isolated target \gls{SE} sources and adaptation to real recording conditions without access to strong labels consisting of the isolated \gls{SE} signals. We perform preliminary experiments in these directions using real mixtures taken from the \noteII{\gls{FSD}}~\cite{fonseca2020fsd50k}.

The remainder of the paper is as follows. We review related works in Section \ref{sec:related}. The \gls{TSE} problem is formalized while introducing the notations in Section \ref{sec:problem}. We present details of the \gls{TSE} frameworks by introducing the class label, enrollment, and SoundBeam in Section \ref{sec:SoundBeam}. In Section \ref{sec:practical issues}, we discuss how to handle the four practical aspects discussed above. We then provide experimental results comparing the different \gls{TSE} frameworks in Section \ref{sec:expe}. Finally, Section \ref{sec:conclusion} concludes the paper and previews possible future works.

\section{Related works}
\label{sec:related}
\subsection{Source separation}
\gls{TSE} is related to source separation since both problems are regression problems from a mixture signal. Source separation, including speech~\cite{kolbaek2017multitalker,hershey2016deep,luo2019conv}, music~\cite{Cano_2019,Uhlich_icassp17,luo2017_deepclustering,rafii2018overview,kumar18c_interspeech}, and universal sound separation~\cite{kavalerov2019universal,FUSS_wisdom2021s,tzinis2020sudo}, has made tremendous progress with the advent of \glspl{DNN}. We can borrow some of the ideas proposed for source separation directly to design \gls{TSE} systems, such as the model architectures and the training losses.
For example, we derive our implementation of the different \gls{TSE} frameworks from the \gls{ConvTasNet}~\cite{luo2019conv}, which was originally proposed for speech separation.

\noteII{
There is a vast diversity of sounds occurring in our everyday environments. Therefore, universal sound separation systems need to handle a large number of possible \gls{SE} classes, $N$. Here, the number of SE classes, $N$, includes all the SE classes that can occur in any recording we wish to process. Each mixture would typically consist of a smaller number of sound sources, $M$, taken from a subset of the possible \gls{SE} classes. In this paper, we consider all sounds that are from the same \gls{SE} class as a single source in the mixture. 
A system that could handle recordings in various situations, such as street, home, orchestra, train stations, etc., would have a large number of possible SE classes, $N$. However, since all situations do not contain sounds from all SE classes, the number of sources in a mixture, $M$, is generally smaller, i.e., $M < N$.
In our experiments, we use datasets with a number of possible \gls{SE} classes, $N$, of 41 or 61, but a number of sound sources in a mixture, $M$, from two to nine, as described in Section \ref{sec:expe} and in Table \ref{tab:datasets}.}

There are two categories of sound separation systems.
The first type of system uses \textit{as many outputs as the total number of possible \gls{SE} classes}, \noteII{$N$}~\cite{Cano_2019,rafii2018overview,olvera2021foreground,Uhlich_icassp17}.
The separation systems in this category perform both separation and identification and could thus be used directly for \gls{TSE}. These approaches have mainly been investigated for music processing, where the number of possible \gls{SE} classes, \noteII{$N$}, is relatively small, e.g., four instruments ($N=4$) in many studies~\cite{Cano_2019}. It may be challenging to scale these systems to a \noteII{large number of \gls{SE} classes, $N$,} that a \gls{TSE} system would require because the computational complexity would increase significantly~\cite{kumar18c_interspeech}.  

As the second type of sound separation system, \gls{USS} proposes instead to use \textit{as many outputs as the maximum number of sources in a mixture, \noteII{$M$}}, which is typically much smaller than the number of \gls{SE} classes, \noteII{$N$}~\cite{kavalerov2019universal}. Such separation systems can be trained using \gls{PIT}~\cite{kolbaek2017multitalker}. \gls{USS} systems can handle an arbitrarily large number of \gls{SE} classes, $N$, and potentially new classes not encountered during training. However, they are limited by the number of sources in the mixtures, $M$, they can handle. Arguably, this may be a significant issue because the number of \glspl{SE} in our everyday surroundings \noteII{vary significantly and can be relatively large}. Moreover, they separate the mixtures without identifying the sounds.  

\gls{TSE} could be achieved by combining \gls{USS} with \gls{SEC}, i.e., first separating the mixture into all its source signals and then identifying the target \gls{SE} class using \gls{SEC}. \noteNI{Note that some works proposed to combine source separation and \gls{SEC}/\gls{SED} to improve \gls{SED} performance but not to achieve \gls{TSE}~\cite{heittola2011sound,gemmeke2013exemplar,CornellDCASE2020,turpault2020improving,huang2020guided}.}
To the best of our knowledge, combining \gls{USS} with \gls{SEC} has not been proposed explicitly for \gls{TSE}, but we still consider it a natural baseline in our experiments of Section \ref{sec:exp_single}.
Although intuitive, this approach has several drawbacks. First, it inherits from \gls{USS} the requirement of knowing or estimating the number of sources in the mixture, $M$. Second, the \gls{SEC} process must be performed on all separated signals, which may become computationally expensive when dealing with many potential sources. Third, such a cascade combination may not be optimal since source separation may introduce errors or processing artifacts that can impact \gls{SEC} performance. Besides, the separation process is independent of the target \gls{SE} class, although knowing the sound characteristics of the target \gls{SE} class could help improve separation.

\subsection{Target sound extraction}
\label{sec:related_tse}
\noteII{Recently, there has been increased interest in approaches that aim at extracting a target signal in an audio mixture~\cite{zmolikova2019speakerbeam,Wang2019voicefilter,LeeCL19,zhao2018sound,seetharaman2019class,slizovskaia2019end,samuel2020meta,kong2020source,Ochiai2020,gfeller2020oneshot,delcroix2021few,okamoto2021environmental}. These works cover various domains of applications and are implemented with various network architectures. However, they share the same idea of conditioning the extraction process on auxiliary clues to identify a target signal in a mixture.}

The domain of application of \gls{TSE} includes speech~\cite{zmolikova2019speakerbeam,Wang2019voicefilter}, 
music~\cite{LeeCL19,zhao2018sound,seetharaman2019class,slizovskaia2019end,samuel2020meta}, and universal sounds~\cite{kong2020source,Ochiai2020,gfeller2020oneshot,delcroix2021few,okamoto2021environmental}.
Various types of auxiliary clues have been proposed, including enrollment audio samples~\cite{zmolikova2019speakerbeam,Wang2019voicefilter,seetharaman2019class,LeeCL19,gfeller2020oneshot}, class labels~\cite{kong2020source,Ochiai2020,samuel2020meta}, video signals of the target source \cite{zhao2018sound}, and recently even onomatopoeia \cite{okamoto2021environmental}.
We deal with universal sounds, which is challenging because it implies dealing with a much larger number of \gls{SE} classes, $N$, than for music (i.e., in this paper, we use up to 61 \gls{SE} classes) and with a greater variety of sounds than speech signals. 

This work focuses on approaches based on enrollment and class labels.  Enrollment-based \gls{TSE} was introduced first for target speech and music  extraction~\cite{zmolikova2017speaker,LeeCL19}, where the enrollment sample consists of an utterance from the target speaker or an audio query of a target instrument. Similar ideas have recently been applied to sound extraction~\cite{gfeller2020oneshot}. 
Meanwhile, class label-based \gls{TSE} was introduced in concurrent works~\cite{kong2020source,Ochiai2020}.

We reproduce systems based on the concepts of prior enrollment- and class label-based \gls{TSE} approaches~\cite{kong2020source,Ochiai2020,gfeller2020oneshot}  using the same formalization to allow a fair comparison between them. 
The main difference between this paper and previous works~\cite{kong2020source,gfeller2020oneshot} is that we base our experiments on a dataset that contains isolated \glspl{SE} annotated with class labels, whereas they did not assume that class labels were available. 
However, we could also use similar schemes to previous works \cite{kong2020source,gfeller2020oneshot} to train SoundBeam on datasets without class labels annotations.
Furthermore, the model architecture also differs, i.e., our study is based on \gls{ConvTasNet} (as in e.g.,~\cite{Ochiai2020}),  while the above works~\cite{kong2020source,gfeller2020oneshot} are based on U-Net.

\subsection{Training on real mixtures}
\label{ssec:related_real}
Training a system with real mixtures requires modifying the classical supervised learning schemes since isolated reference sources are unavailable. There have been several proposals to train separation systems from real mixtures, including training on mixtures-of-mixtures \cite{wisdom2020unsupervised} and using losses that do not require clean sources, such as a \gls{GAN}~\cite{zhang2017weakly,stoller2018adversarial} or an \gls{SEC}-based loss \cite{pishdadian2020finding}.
Several \gls{TSE} systems have been trained with a similar principle as mixtures-of-mixtures~\cite{gfeller2020oneshot,kong2020source,zhao2018sound,gao2019co}. 
For example, one work~\cite{kong2020source} used \gls{SED} to detect isolated \gls{SE} in real mixtures, while another \cite{gfeller2020oneshot} showed that the enrollment-based \gls{TSE} could also be trained even when the audio query consists of sound mixtures. 

Although mixtures-of-mixtures approaches permit training with real mixtures, they still generate artificial recordings since unrelated audio signals are mixed. An alternative consists of retraining directly on real mixtures using a weakly supervised loss based on~\gls{SEC}~\cite{pishdadian2020finding}. We adopt this scheme originally proposed for training separation systems with a small number of \gls{SE} classes \noteNI{(i.e., $N=5$ in \cite{pishdadian2020finding})} and apply it to adapt a \gls{TSE} system on real data with a large number of classes \noteNI{(i.e., $N=61$, which corresponds to the number of \gls{SE} class encountered during training in our experiments, as described in Section \ref{ssec:expe_data})}.

\section{Target sound extraction problem formulation}
\label{sec:problem}

We consider the problem of extracting sounds of one or multiple target \gls{SE} classes from a mixture of \glspl{SE} captured with a single microphone.
The observed mixture signal is given as,
\begin{align}
    \y = \sum_{n=1}^N \x^n + \b,
    \label{eq:mixture}
\end{align}
where $\y \in \mathbb{R}^{T}$, $\x^n \in \mathbb{R}^{T}$, and $\b \in \mathbb{R}^{T}$ are the observed single-channel mixture, the source signal from the $n$-th \gls{SE} class, and the background noise, respectively. 
\noteII{The background noise, $\b$, includes ambient noise and could also potentially include sounds from SEs that are not defined in the SE classes of the system. }
$T$ is the signal duration. 

We assume $\x^n=\mathbf{0}$ when no source from the $n$-th \gls{SE} class is active, where $\mathbf{0}$ denotes a vector of all zeros. Later, we refer to this case as an \textit{inactive \gls{SE} class}. Note that $\x^n$ may consist of multiple \gls{SE} sources from the same \gls{SE} class. For example, if the mixture includes barking sounds from several dogs, the source signal $\x^n$ for the barking \gls{SE} class will consist of a mixture of all barking sounds.
\noteII{In Eq.~\eqref{eq:mixture}, the summation is over all possible \gls{SE} classes, $N$, including target and non-target ones. Since, in practice, many \gls{SE} classes are often inactive, the actual number of sources in a mixture, $M$, is usually smaller than the total number of \gls{SE} classes, $N$.}

\subsection{Single target extraction}
\noteII{
In this paper, we mostly deal with \gls{TSE} of a single target \gls{SE} class. In this case, the goal of \gls{TSE} is to estimate the target signal, $\x^{s}$, where $s$ is the index of the target \gls{SE} class. If the target \gls{SE} class is inactive, \gls{TSE} should estimate a zero signal, i.e., $\x^{s} = \mathbf{0}$.}

\noteII{\gls{TSE} requires clues to indicate the target \gls{SE} classe.
In the following, we consider two types of clues, i.e., class labels and enrollment audio samples.
Target class labels can be represented as 1-hot vectors, $\o=[o^s_1, \ldots, o^s_N]^T$, where
\begin{align}
    o^s_n =\left\{ \begin{array}{ll}
         1 & \mbox{if $n = s$},\\
         0 & \mbox{if $n \neq s$}. 
         \end{array} \right. 
         \label{eq:1-hot}
\end{align}
A class label-based \gls{TSE} system estimates thus the target source given the 1-hot vector as,
\begin{align}
    \xhat = \TSE(\y, \o),
\end{align}
where $\xhat \in \mathbb{R}^{T}$ is an estimate of the target source and $\TSE(\cdot)$ represents a \gls{TSE} system.}

\noteII{
Alternatively, we can also use an enrollment audio sample of the target \gls{SE} class, } \noteI{ $\a \in \mathbb{R}^{T_a}$}\noteII{, as a clue, where } \noteI{ $T_a$} \noteII{ is the duration of the enrollment sample.
An enrollment-based \gls{TSE} system performs thus the following, 
\begin{align}
    \xhat = \TSE(\y, \a).
\end{align}
}
\subsection{Multi-target extraction}
\noteII{
We can generalize \gls{TSE} to the simultaneous extraction of sounds from multiple target \gls{SE} classes\cite{Ochiai2020}.
We denote by $S=\{s_j\}_{j=1}^J$ the set of indexes of the target \gls{SE} classes that we want to extract, where} \noteI{$J$ is the number of target \gls{SE} classes, which can differ for each processed mixture. For example, if we want to extract sounds from two classes simultaneously ($J=2$), such as both piano and guitar, $S=\{s_1=n_{\text{piano}}, s_2=n_{\text{guitar}}\}$, where $n_{\text{piano}}$ and $n_{\text{guitar}}$ are the indexes of the piano and guitar classes, respectively. }\noteII{
The goal of \gls{TSE} becomes estimating the target signal $\x^{S}$, which consists of the sum of the signals from all target \gls{SE} classes: 
\begin{align}
    \x^{S} = \sum_{n \in S} \x^n .
\end{align}
Note that if all target \glspl{SE} are inactive, \gls{TSE} should estimate a zero signal, i.e., $\x^{S} = \mathbf{0}$.}

\noteII{When using class labels clues, we can generalize the 1-hot vector of Eq. \eqref{eq:1-hot} to a n-hot vector, $\oS=[o^S_1, \ldots, o^S_N]^T$, where
\begin{align}
    o^S_n =\left\{ \begin{array}{ll}
         1 & \mbox{if $n \in S$};\\
         0 & \mbox{if $n \notin S$}. 
         \end{array} \right. 
         \label{eq:n-hot}
\end{align}
A class label-based multi-target \gls{TSE} system estimates the mixture of the target sources as,
\begin{align}
    \xhatS = \TSE(\y, \oS),
\end{align}
where $\xhatS$ is an estimate of the target signal $\xS$.}

\noteII{When using enrollment audio samples as clues, we consider here that we have separate enrollment samples for each of the target \gls{SE} classes. An enrollment-based multi-target \gls{TSE} system performs thus the following,
\begin{align}
    \xhatS = \TSE(\y, \{\a\}_{s\in S}),
\end{align}
where $\{\a\}_{s\in S}=\{\mathbf{a}_{s_1}, \ldots, \mathbf{a}_{s_J} \}$ is the set of separate enrollment audio samples from each target \gls{SE} class. }

\noteII{
In most parts of the paper, we use the notations for single target extraction to simplify the explanations. We discuss the multi-target extraction in Section \ref{sec:multiclass}.}

\section{SoundBeam framework for TSE}
\label{SoundBeam}
\begin{figure*}[tb]

    \centering
    \includegraphics[width=0.95\textwidth]{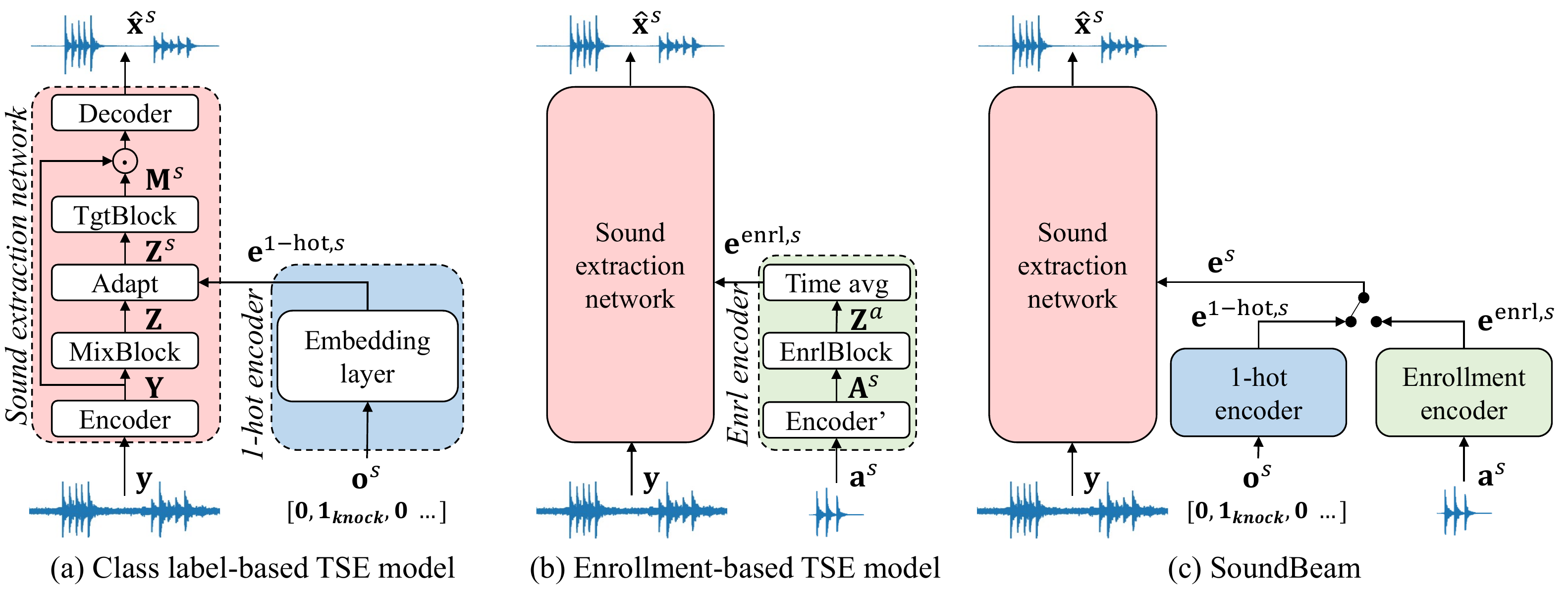}
    \caption{Diagrams of three different \gls{TSE} frameworks.}
    \label{fig:method}
\end{figure*}

In this section, we introduce the generic framework for \gls{NN}-based \gls{TSE} from which we derive SoundBeam. The \gls{TSE} models share a common structure and are composed of two modules, i.e., (1) a sound extraction \gls{NN} and (2) a clue encoder that computes the target \gls{SE} embeddings. 
Figure \ref{fig:method} schematically overviews the \gls{TSE} frameworks for (a) class label-based \gls{TSE}, (b) enrollment-based \gls{TSE}, and (c) SoundBeam.

\label{sec:SoundBeam}

\subsection{Sound extraction NN}
The sound extraction \gls{NN} estimates the target signal, $\xs$, from the mixture, $\y$, and \noteI{a $D$-dimensional target embedding vector}, \noteI{$\es \in \mathbb{R}^{D}$}, as,
\begin{equation}
    \xhat = f(\y,\es),
\end{equation}
where $f(\cdot)$ is the sound extraction \gls{NN}.
The embedding vector, $\es$, represents the characteristics of the target \gls{SE} class. It is computed with the clue encoder derived from the class label or enrollment clues described in section \ref{ssec:embedding}. 

There are many possible ways to implement the extraction \gls{NN}. We borrow the implementation from \gls{ConvTasNet} because it has been widely used for speech separation~\cite{luo2019conv} and target speech extraction~\cite{delcroix2020improving}. However, our derivation is general enough to be extended to other network architectures such as U-Net or to other systems using the \gls{STFT}/\gls{iSTFT} as encoder/decoder layers.
The detailed processing is as follows.

The time domain input signal, $\y$, is processed by a trainable encoder layer as,
\begin{equation}
    \Y = \enc(\y), \label{eq:encoder layer}
\end{equation}
where $\Y \in \mathbb{R}^{D \times T'}$ is the encoded mixture representation, \noteI{$T'$ is the number of time frames}, and $\enc(\cdot)$ is the encoder layer, which consists of a 1-D convolution layer~\cite{luo2019conv}. 
The estimated target signal, $\xhat$, is obtained as,
\begin{align}
    \xhat &= \dec(\Ms \odot \Y),
\end{align}
where $\Ms$ is a target mask, 
$\odot$ is the element-wise product,
and $\dec(\cdot)$ is a decoder layer that maps the output representation of the extracted signal back to the time domain.

Compared to \gls{ConvTasNet}, \gls{TSE} only requires computing the mask for the target \gls{SE} class instead of one for each source. 
The target mask, $\Ms$, is computed as follows,
\begin{align}
    \Z &= \extmix(\Y), \\
    \Zs &= \adapt(\Z, \es), \\
    \Ms &= \exttgt(\Zs),
\end{align}
where $\Z\in \mathbb{R}^{D\times T'}$ and
$\Zs\in \mathbb{R}^{D\times T'}$ are the internal representation of the mixture and the target, respectively.
$\extmix(\cdot)$ is the lower part of the sound extraction \gls{NN}, which transforms $\Y$ into a general internal representation of the mixture independent of the target \gls{SE} class. 
$\exttgt(\cdot)$ is the upper part of the sound extraction \gls{NN}, which computes the target mask, $\Ms$, that specifically extracts the sounds from the target \gls{SE} classes \noteNI{(note that in \cite{delcroix2021few}, $\extmix(\cdot)$ and $\exttgt(\cdot)$ are referred to as $\ExtractBlockI(\cdot)$ and $\ExtractBlockII(\cdot)$, respectively)}.
$\adapt(\cdot)$ is an adaptation layer. $\extmix(\cdot)$ and $\exttgt(\cdot)$ are implemented as stacks of 1-D convolutional blocks~\cite{luo2019conv}.

One of the key differences from the original \gls{ConvTasNet} is the adaptation layer, which combines the internal representation of the mixture, $\Z$, with the target \gls{SE} embedding, $\es$. It plays the crucial role of conditioning the extraction process on the target \gls{SE} classes. Various types of adaptation layers have been proposed in the context of target speech extraction, including factorized~\cite{zmolikova2019speakerbeam}, addition/concatenation~\cite{Wang2019voicefilter}, multiplication~\cite{delcroix2020improving}, attention~\cite{xiao2019single}, and \gls{FiLM}~\cite{gfeller2020oneshot} layers. Here, we use the simple yet sufficiently powerful~\cite{zmolikova2019speakerbeam} element-wise multiplication layer as,
\begin{equation}
\zs_t = \z_t \odot \es,
\end{equation}
where \noteI{$\zs_t \in \mathbb{R}^{D} $ and $\z_t \in \mathbb{R}^{D}$} are the $t$-th frame of $\Z$ and $\Zs$, respectively.

\subsection{Clue encoder}
\label{ssec:embedding}
The different \gls{TSE} frameworks differ by the clue encoders they use.
The clue encoder computes the \gls{SE} embedding vectors from the class label or enrollment clues. Below, we describe these two configurations and a mixed encoder, which SoundBeam employs.

\subsubsection{Class label encoder}
\label{sec:1-hot_encoder}
The class label encoder consists of an embedding layer that converts the 1-hot vector, $\o$, defined in Eq. (\ref{eq:n-hot}) into a continuous representation as,
\begin{equation}
    \eo = \W \o,
    \label{eq:1hot embedding}
\end{equation}
where \noteI{$\W = [ \mathbf{e}^{\text{class}, 1}, \ldots, \mathbf{e}^{\text{class}, N}]\in \mathbb{R}^{D \times N}$} is an embedding matrix whose columns contain the
embedding vectors of each \gls{SE} class. 

Figure \ref{fig:method}(a) shows a diagram of the class label-based \gls{TSE} model similar to previous works \cite{Ochiai2020,kong2020source}. By learning the class label encoder and the sound extraction \gls{NN} jointly, we can optimize the \gls{SE} embedding vectors directly for the TSE task. However, the \gls{SE} classes that the method can handle are fixed by the embedding matrix, $W$, and limited to the $N$ \textit{``seen’’ \gls{SE} classes} encountered during training. Consequently, we cannot directly use it for the extraction of new \gls{SE} classes.

\subsubsection{Enrollment encoder}
\label{sec:enrl_encoder}
The enrollment encoder computes the target \gls{SE} embedding vector from an enrollment audio sample of the target \gls{SE} class, $\a$. Here, we use a sequence summary \gls{NN}~\cite{vesely2016sequence,zmolikova2019speakerbeam} to convert the variable-length enrollment, $\a$, to a vector of fixed dimension as,
\begin{equation}
    \ea = g(\a).
\end{equation}
$g(\cdot)$ is an enrollment encoder \gls{NN} implemented as,
\begin{align}
    \A &= \enc(\a) \\
    \Z^a &= \enrlnet(\A) \\
    \ea &= \frac{1}{T_a'} \sum_t \z_t^a, \label{eq:seqsum}
\end{align}
where \noteI{$\enc(\cdot)$ is an encoder layer similar to that in Eq. (\ref{eq:encoder layer}) but with different parameters}, $\enrlnet(\cdot)$ consists of a stack of 1-D convolution blocks, $\A \in \mathbb{R}^{D \times T_a'}$ and $\Z^a \in \mathbb{R}^{D \times T_a'}$ are internal representations of the enrollment, $\z_t^a \in \mathbb{R}^{D}$ is the $t$-th time frame of $\Z^a$, and $T_a'$ is the number of time frames. 
The last layer performs average pooling over the time dimension.

Figure \ref{fig:method}(b) shows a diagram of enrollment-based \gls{TSE} similar to previous works~\cite{seetharaman2019class,gfeller2020oneshot}.
Compared to the class label-based \gls{TSE}, the enrollment-based approach does not explicitly rely on well-defined class labels but instead identifies and extracts sounds in the mixture based on their similarity to the enrollment. Consequently, the enrollment-based model can generalize to \noteII{a new target \gls{SE} class} as long as an enrollment audio sample of the new class can be collected and the model can be trained with a sufficient variety of \gls{SE} classes.
On the other hand, the embedding vector, $\ea$, is not directly optimized for the target \gls{SE} class, which may result in lower performance compared with the class label-based model when extracting sounds from seen \gls{SE} classes.

\subsubsection{Mixed encoder of SoundBeam}
\label{sec:mixed_encoder}
Both class label and enrollment models use an embedding vector to characterize the target \gls{SE} class. When trained independently, the embedding vectors of both models are mapped to different embedding spaces. However, we can enforce mapping them to a joint embedding space by designing a model that shares the sound extraction \gls{NN} and alternates between the class label and enrollment encoders during training. 
\noteIII{The sound extraction network would thus be trained to output the same extracted signal when conditioning on embedding vectors derived from either the class labels or enrollment samples. One way to achieve this is if the class label and enrollment encoders provide similar representations for \glspl{SE} of the same class, i.e., the embedding vectors derived from the class labels and enrollment samples are similar and mapped to the same region of the embedding space. Note that we could also include explicit constraints to further enforce learning a joint embedding space\cite{delcroix2021few}, but in the experiments of this paper, such an additional constraint was not necessary.}

At test time, 
we can use either the class label or the enrollment encoder to perform the extraction.
 Figure \ref{fig:method}(c) is a diagram of SoundBeam, which uses the mixed encoder. The switch indicates that we can use either the class label or the enrollment encoder.

SoundBeam offers several advantages. First, the alternate training scheme, described in Section \ref{sec:training}, acts as multi-task training, which may lead to better performance. Moreover, we can employ at test time the class label encoder to use optimal target \gls{SE} embeddings for seen \gls{SE} classes or the enrollment encoder to handle new classes. Finally, we can derive the few-shot adaptation scheme described in section \ref{sec:new_SE}, which enables us to learn optimal embeddings for new \gls{SE} classes with a few enrollments. 

\subsection{Training TSE systems}
\label{sec:training}

This section introduces the fully supervised training scheme of the \gls{TSE} systems, where we assume that the sound mixture, $\y$, the target sound signal, $\x^s$, and the class label or enrollment clues are available. We learn the sound extraction \gls{NN} and the clue encoder jointly to ensure that the embedding vectors capture the information needed for the TSE task.

We use the negative thresholded \gls{SDR}~\cite{wisdom2020unsupervised} as extraction loss,
\begin{align}
\mathcal{L}^{\text{ext}} (\xhat, \x^s )& =  - 10 \log_{10} \left( \frac{\| \x^s \|^{2}}{\| \x^s - \xhat \|^{2} + \tau \| \x^s \|^2} \right), \label{eq:ext_loss} \\
& \Leftrightarrow  10 \log_{10} \left( \| \x^s - \xhat \|^{2} + \tau \| \x^s \|^2 \right) \label{eq:ext_loss_den}
\end{align}
where $\tau$ is a threshold that limits the upper value of the loss, thus preventing the low error/distortion training samples from dominating the gradient. \noteII{Here, we follow the prior work~\cite{wisdom2020unsupervised} and fix in all experiments $\tau = 10^{-\frac{\eta}{10}}$, where $\eta = 30$ dB is a soft threshold value following}.

To train SoundBeam, we sum the extraction loss obtained with the class label and enrollment encoders as,
\begin{align}
    \mathcal{L}^{\text{SoundBeam}} = & \alpha \mathcal{L}^{\text{ext}}\left(\xhat=f\left(\y, \eo\right), \x^s \right) \nonumber\\
    &+ (1-\alpha) \mathcal{L}^{\text{ext}}\left(\xhat=f\left(\y, \ea\right), \x^s \right),  
    \label{eq:mt_loss}
\end{align}
where $\alpha$ is a multi-task weight that we fix to $\alpha=0.5$ in all experiments.
This loss is computed by propagating each training sample twice by alternating the clue encoders. \noteI{In other words, for each mini-batch, we first propagate with the class labels encoder ($\xhat=f\left(\y, \eo\right)$) and then with the enrollment encoder ($\xhat=f\left(\y, \ea\right)$) while accumulating the gradients, so that each training sample is extracted with both clues.} This enables multi-task learning, which can have a beneficial effect on performance.
More importantly, SoundBeam learns a common embedding space for class label-based and enrollment-based TSE by sharing the same sound extraction \gls{NN} and training simultaneously with both the class label and enrollment encoders.

\section{Practical issues}
\label{sec:practical issues}

\subsection{Extraction of new SE classes with few-shot adaptation}
\label{sec:new_SE}
It is a challenging task to create a universal \gls{TSE} system that can cover all
\gls{SE} classes at its initial deployment since there is a wide variety of possible \glspl{SE} that occur in our daily environments. In addition, the target \gls{SE} classes may depend on the environments and applications.
We believe that an essential property of a \gls{TSE} system is to extend its capabilities flexibly and be able to continuously learn how to extract new target \gls{SE} classes given a few audio samples of the new classes.

The class label model cannot handle new \gls{SE} classes as discussed in Section \ref{sec:1-hot_encoder}.
The enrollment model performs extraction based on sound similarities between the characteristics of the sounds in the mixture and the enrollment. Consequently, it can naturally generalize to new \gls{SE} classes if we can record enrollment audio samples for these classes. However, the embedding vectors are not optimized directly for extracting the new \gls{SE} classes.
Using SoundBeam, we can combine the advantages of both approaches and learn to extract sounds from new \gls{SE} classes using a few enrollments (few-shot adaptation). We describe the procedure for this below.

Let \noteI{$\tilde{s}=N+1$} be the index of a new \gls{SE} class.
\noteNI{Here, we describe the addition of a single new \gls{SE} class, but we can equivalently add multiple new \gls{SE} classes simultaneously.}
We assume that we can collect $K$ enrollments, \noteI{$\{\astar_1, \ldots, \astar_K\}$}, from that new class.
First, we compute an average embedding, \noteI{$\estar$}, for the new \gls{SE} class using the enrollment encoder as,
\begin{align}
    \estar = \frac{1}{K} \sum_{k=1}^K g(\astar_k ).
    \label{eq:avg_emb}
\end{align}
Because SoundBeam learns a common embedding space for the embedding vectors obtained with the class label and enrollment encoders, this average embedding can be used as a new entry to the embedding matrix of the class label encoder as,
\noteI{
\begin{align}
    \tilde{\W} = [\W, \estar]\in \mathbb{R}^{D \times (N+1)},
\end{align}}
where \noteI{$\tilde{\W} $} is a new embedding matrix.
This process would already enable extracting sounds from the new \gls{SE} class with the class label encoder, but the new embedding vector is not yet optimized for the \gls{TSE} purpose. 

Consequently, we propose to fine-tune the new embedding vector, $\estar$, which we call \textit{few-shot adaptation to new \gls{SE} classes}. First, we generate adaptation data by mixing the $K$ enrollments with \gls{SE} signals from the original training data.
We then retrain the new embedding vector, $\estar$, using the adaptation data following a similar training procedure as in Section \ref{sec:training}.
Note that we can optimize the embedding vector of the new \gls{SE} class while preserving the performance on the
original \gls{SE} classes by fixing all model parameters but the new embedding vector during the fine-tuning process.
This simple approach allows learning new \gls{SE} classes with few samples while avoiding catastrophic forgetting~\cite{goodfellow2013empirical}, making it suitable for continuous learning. 

\noteII{Note that we assume that we could record a few enrollment audio samples of the new \gls{SE} classes. This implies being able to record isolated \glspl{SE} of the new class, which may be challenging in real-life recordings where many sounds often overlap. Since our proposed few-shot adaptation requires only a small number of enrollment samples, we assume that the user could carefully record sounds of the  target \gls{SE} class or select portions of a recording where the target \gls{SE} class is clearly dominant. Extension of the approach for noisy enrollments will be part of our future works.}
\subsection{Handling of inactive target SE classes}
\label{sec:inactive}
Dealing with inactive target \gls{SE} classes is another important issue of \gls{TSE} systems.
For example, when the target \gls{SE} class is dog barking, but there are no barking sounds in the mixture, the output should be silent, i.e., $\x^s=\mathbf{0}$.
However, it remains unclear how well \gls{TSE} systems can learn this behavior since most studies focused only on active \gls{SE} classes. For example, one proposed \gls{TSE} system~\cite{kong2020source} was trained with data that included inactive classes, but the system was not evaluated for inactive class extraction. More recently, another approach \cite{borsdorf21_interspeech} proposed evaluating the extraction for inactive speakers but not for \glspl{SE}. 

The problem of extraction of inactive classes is related to the problem of separation with an unknown number of speakers, where the number of sources in the mixture can be smaller than the number of outputs of the separation system~\cite{FUSS_wisdom2021s}. The separation systems address this problem by outputting zero signals for the inactive sources, which is similar to the inactive class problem for \gls{TSE}. 
The extraction loss of Eq. (\ref{eq:ext_loss}) is ill-defined when $\x^s=\mathbf{0}$.
However, we can define an extraction loss for the inactive \gls{SE} cases as~\cite{FUSS_wisdom2021s},
\begin{equation}
\mathcal{L}^{\text{inactive}} (\xhat, \y )=10 \log_{10} \left( \|  \xhat \|^{2} + \tau^{\text{inactive}} \| \y \|^2  \right),
\end{equation}
where $\tau^{\text{inactive}}$ is a soft threshold set at $\tau^{\text{inactive} } = 10^{-2}$.
This loss consists of the denominator term of Eq. (\ref{eq:ext_loss}) as shown in Eq. (\ref{eq:ext_loss_den}), with a different setting for the soft threshold (i.e., $\x^s$ replaced by $\y$). 

During training, following a previous study~\cite{kong2020source}, we include 10 \% of \gls{IS} to learn how to handle inactive \gls{SE} classes. 
The training loss thus becomes,
\begin{align}
    \mathcal{L} (\xhat, \x^s, \y )= \left\{ \begin{array}{cc}
        \mathcal{L}^{\text{active}} (\xhat, \x^s ) &  \text{if } \x^s \neq \mathbf{0},\\
        \mathcal{L}^{\text{inactive}} (\xhat, \y ) & \text{if } \x^s = \mathbf{0},
    \end{array} \right.
    \label{eq:active&inactive_loss}
\end{align}
where $\mathcal{L}^{\text{active}} (\xhat, \x^s )$ is the active loss defined in Eqs. (\ref{eq:ext_loss}) and (\ref{eq:mt_loss}). For SoundBeam, $\mathcal{L}^{\text{inactive}} (\xhat, \y )$ is computed with the class label and enrollment encoders, as for the active cases. 

\noteII{Note that we opted for a scale-dependent extraction loss for $\mathcal{L}^{\text{active}}$ as shown in Eq.(\ref{eq:ext_loss}) instead of the scale-independent version widely used for source separation~\cite{luo2019conv} because we believe that the scale of the output signal may matter in practical applications to detect inactive target \gls{SE} classes. For example, we can evaluate how well the system could internally detect active/inactive target classes by looking at the attenuation from the mixture, which would not be possible when using a scale-independent loss as the system could choose to output signals with, e.g., very low energy.}


If the \gls{TSE} system could output zero signals for inactive \gls{SE} classes, it would be possible to detect whether a class is active by considering the attenuation ratio between the mixture and the extracted signal, $\mathcal{A}^{\text{mixture}} (\xhat, \y)$, as,
\begin{align}
    \mathcal{A}^{\text{mixture}} (\xhat, \y)= - 10 \log_{10} \left( \frac{\| \y \|^{2}}{\| \xhat \|^{2}} \right). 
    \label{eq:att_mix}
\end{align}
Consequently, an \gls{SE} class would be considered inactive if the ratio were smaller than a threshold.
We use this approach to evaluate the different \gls{TSE} systems in cases of inactive \gls{SE} classes.

\subsection{Simultaneous multi-target extraction}
\label{sec:multiclass}
For some applications, the user may want to hear sounds from multiple target \gls{SE} classes, e.g., both sirens and klaxon sounds, when walking in the street.
We can permit this by independently performing \gls{TSE} for each target \gls{SE} class and then summing up the extracted signals. However, this naive approach would increase the computational complexity by the number of target \gls{SE} classes. We propose instead learning to simultaneously extract a mixture of target \gls{SE} classes~\cite{Ochiai2020}.

For simultaneous multi-target  extraction, the embedding vector should characterize all target \gls{SE} classes. For class label-based \gls{TSE}, we can replace the 1-hot vector, $\o$, with an n-hot vector, $\oS$, that represents multiple \gls{SE} classes as defined in Eq. (\ref{eq:n-hot}).
The class label encoder of Section \ref{sec:1-hot_encoder} can naturally generalize to accept n-hot vectors as,
\begin{equation}
    \eoS = \W \oS = \sum_{s \in S} \W \o.
    \label{eq:n-hot-embedding}
\end{equation}
Equation (\ref{eq:n-hot-embedding}) shows that the multi-target embedding vector, $\eoS$, is simply the sum of the embedding vectors of all target \gls{SE} classes in $S$.

Similarly, for the enrollment-based \gls{TSE}, we define the multi-target embedding vector as the summation of the embedding vectors obtained from the individual enrollments of each class as,
\begin{equation}
    \eaS = \sum_{s \in S} \ea = \sum_{s \in S} g(\a ),
\end{equation}
where $\a$ consists of isolated recordings from each target \gls{SE} class. 

We learn simultaneous extraction by including multi-target training samples, performing extraction with the above multi-target embeddings, and computing the loss with the multi-target signal, $\x^S$, instead of the single-target, $\x^s$, in Eq. (\ref{eq:ext_loss}). During training, we randomly select the number of target \gls{SE} classes so that the same system can learn how to extract various numbers of classes.

\subsection{Challenges faced with real recordings}
\label{sec:realmixtures}
Finally, we discuss the challenges of applying \gls{TSE} to real recordings.
Training a \gls{TSE} system with the supervised loss of Eq. (\ref{eq:ext_loss}) requires access to the target source signal, $\x^s$.
This requirement implies that we can only train with simulated mixtures since the source signals are usually unavailable for real recordings. However, it is challenging to create \noteII{a large number of} realistic mixtures of \gls{SE} sounds by simulation due to, for example, the difficulty of correctly recording isolated \gls{SE} sounds, the diversity of acoustic scenes, sound occurrence patterns, and duration.
Therefore, there may be a significant mismatch between the training and testing conditions, which could impede extraction performance when processing real recordings.

There are various ways to tackle this problem, as discussed in subsection \ref{ssec:related_real}. 
Here, we hypothesize that the classification performance of a \gls{SEC} system should improve the better we extract \glspl{SE}.
Therefore, we explore retraining a \gls{TSE} system on real recordings using a pre-trained \gls{SEC} model to compute a weakly supervised training loss, similar to a previous work~\cite{pishdadian2020finding}, instead of the supervised loss of Eq. (\ref{eq:ext_loss}). 
The \gls{SEC} loss is computed on the output of the \gls{TSE} system as,
\begin{equation}
   \mathcal{L}^{\text{SEC}} (\xhat, s )= \CE \left( c( \xhat), s  \right),
\end{equation}
where $\CE(\cdot)$ is the binary cross-entropy and $c(\cdot)$ is an \gls{SEC} model with fixed parameters~\cite{pishdadian2020finding}.
The \gls{SEC} loss can be computed without clean reference signals as long as the \gls{SE} class labels, $s$, are available (i.e., weak labels). 

A similar loss has been proposed to train sound separation models~\cite{pishdadian2020finding}. However, compared with separation, \gls{TSE} uses clues about the target \gls{SE} class as an auxiliary input. Therefore, the system can easily exploit the information about the target \gls{SE} to maximize the classification loss independently of the input mixture. We avoid this issue by pre-training the system using the fully supervised loss of Eq. (\ref{eq:ext_loss}) and only retraining the lower layers of the \gls{TSE} system, which are not exposed to the target \gls{SE} clues.

\section{Experiments}
\label{sec:expe}
We evaluate the class label, enrollment, and SoundBeam models,
and compare their performance for handling new \gls{SE} classes,
inactive classes, multi-target extraction, 
and real data.

\subsection{Data}
\label{ssec:expe_data}

\begin{table}[t]
  \caption{Details of different Datasets.}
  \label{tab:datasets}
  \centering
  \begin{tabular}{@{}l@{ }c c c c@{ }c@{ }c@{}}
\toprule
Data set & Nb       & Audio  & No. of   & \multicolumn{3}{c}{No. of mixtures} \\
         & Classes $N$ & Dur. $T$ & sources $M$ & Train & Valid. & Test \\
\midrule
Single target  (41)   & 41 & 6s & 3 & 50k & 10k & 3k \\
Single target  (61)   & 61 & 6s& 3 & 50k & 10k & 3k \\
Multi-target        & 61 & 6s& 3-5 & 50k & 10k & 3k \\
Few-shot adaptation     & 41+20 & 6s & 3 & 3k & 500 & - \\
\midrule
Real mixtures               & 61 & 1-30s &2-9 & 406 & 175 & 727 \\
\gls{SEC}          & 61 & 1-30s  & 1 & 18.8k & - & 3.6k \\
\bottomrule
  \end{tabular}
\end{table}

To compare the effectiveness of the \gls{TSE} frameworks, we created several datasets of simulated and real sound event mixtures based on the \gls{FSD} corpora, including \gls{FSD}-Kaggle 2018~\cite{fonseca2018general} and \gls{FSD}50K~\cite{fonseca2020fsd50k}. In all experiments, we downsampled the sounds to 8 kHz to reduce the computational and memory costs. Table \ref{tab:datasets} summarizes the details of the different datasets.

\subsubsection{Single target dataset} The single-target dataset consists of simulated mixtures.
We created mixtures by mixing several \glspl{SE}  selected randomly from different \gls{SE} classes ($N=41$ or $N=61$, depending on the experiment). The \glspl{SE} include human sounds, object sounds, and musical instruments. 
\noteII{We selected the 41 \gls{SE} classes with the most training samples in the FSD corpora. On average, each \gls{SE} class has 220 samples in the training set. We added 20 additional \gls{SE} classes to generate the 61 \gls{SE} class dataset. We chose the additional \gls{SE} classes that were relatively well represented in the \gls{FSD} corpora. However, they had fewer samples in the training data, i.e., on average, 47 samples per \gls{SE} class.}
We generated six-second-long mixtures ($T=6$ secs) using the Scaper toolkit~\cite{salamon2017scaper} by inserting isolated \glspl{SE} at random time positions on top of the background noise. The isolated \glspl{SE} consisted of signals of 2 to 5 secs randomly selected from the \gls{FSD} corpora. 
The background noise consisted of stationary noise from the  REVERB challenge corpus~\cite{kinoshita2016summary} mixed at a \gls{SNR} between 15 and 25 dB. The mixtures were composed of three \glspl{SE} ($M=3$).
For the enrollment-based experiments, we randomly selected a sample from the target \gls{SE} class that differs from the target sound in the mixture.

\subsubsection{Multi-target dataset}
\label{ssec:multi-dataset}
For the multi-target \gls{TSE} experiments, we created a dataset of mixtures created similarly to the single-target dataset but with three to five \glspl{SE} ($M=3\sim 5$) per mixture.

\subsubsection{Few-shot adaptation dataset} For the few-shot adaptation experiments, we consider the 41 classes of the single-target dataset as seen classes and the remaining 20 classes as new.  We sampled $K$ enrollments from each new \gls{SE} class, with $K=1, \ldots, 15$. We used these enrollments to compute the average embedding of Eq. (\ref{eq:avg_emb}) and to generate adaptation data. 
We created the adaptation data by mixing one enrollment of the new \gls{SE} classes with two \gls{SE} training samples from the 41 seen classes, using a similar simulation procedure as above. The total number of adaptation data consisted of 3000 mixtures, covering all 20 new \gls{SE} classes. \noteNI{Here, the adaptation data contains all 20 new \gls{SE} classes because we perform the adaptation of multiple new \gls{SE} classes simultaneously. However, we could also adapt the models for each new class independently.} We generated six trials for the adaptation experiments by varying the random seed for the sampling enrollments. We used the test set of the single-target class datasets with 61 classes for evaluation.

\subsubsection{Real mixture dataset} In addition to the simulated mixtures, we also used a small dataset of real mixtures that consists of recordings from the \gls{FSD}50K that contained two or more labeled \gls{SE} classes. The recording and mixing conditions vary considerably compared to the simulated mixtures. For example, most mixtures had two to three classes, but some contained up to nine, and the duration varied between 1 to 30 seconds.
\noteII{Moreover, although the recordings include multiple \gls{SE} classes, there is no control over the actual temporal overlap. Therefore, we can reasonably expect that some of the real recordings consist of partially or non-overlapping \glspl{SE}.}

\subsubsection{SEC dataset} Finally, we used the original training data from both the \gls{FSD}-Kaggle and the \gls{FSD}50K corpora to retrain the \gls{SEC} system.

\subsection{Experimental settings}
We used the Asteroid toolkit for all experiments~\cite{Pariente2020Asteroid}.

\subsubsection{TSE models}
The configuration of the different \gls{TSE} models followed \gls{ConvTasNet}~\cite{luo2019conv}. In the explanations below, we follow the notations of original \gls{ConvTasNet}~\cite{luo2019conv}, except for the encoded features dimension $D$. In particular, the encoder and decoder layers consisted of 1-D convolution and deconvolution layers that operated on segments of $L=20$ taps with 50\% overlap. The dimension of the encoded features was $D=256$.  $\extmix$, $\enrlnet$ and $\exttgt$
in Figure~\ref{fig:method} consisted of stacked dilated 1-D  convolutional blocks with $H=512$ channels, kernel size of $P=3$, and $B=256$ bottleneck channels.  $\extmix$ and $\enrlnet$ consisted of a single stack ($R=1$), while in $\exttgt$ the stack was repeated seven times ($R=7$). We used $X=8$ convolutional blocks per stack for $\extmix$ and $\exttgt$, and $X=4$ for $\enrlnet$.

We trained all models using the Adam optimizer~\cite{kingma2015adam} with a learning rate of $10^{-4}$, a batch size of 8, and a maximum of 200 epochs. 
In all experiments, we used the models achieving the lowest cross-validation loss value. 

For the few-shot adaptation experiments, we fixed all network parameters except the embedding layer and trained for 50 epochs with a learning rate of $10^{-3}$. 

For retraining with the weakly supervised \gls{SEC} loss, we fixed all parameters except for $\extmix$ (i.e., the part of the extraction \gls{NN} not exposed to the auxiliary inputs). This scheme prevents the network from exploiting the embedding vectors to reduce the \gls{SEC} loss artificially.

\subsubsection{USS model}
We compared the \gls{TSE} models with a \gls{USS} system based on \gls{ConvTasNet} with three outputs. We used a similar network configuration for the separation system to that of SoundBeam and trained it using \gls{PIT}.

To realize \gls{TSE}, we chose from among the outputs of the separation the one output with the highest posterior probability for the target \gls{SE} class. We computed the posteriors with the \gls{SEC} model described below.
\subsubsection{SEC model}
We used the publicly available \gls{PANN} model~\cite{kong2020panns} for our experiments with \gls{SEC}. We retrained the model for \gls{SEC} on the \gls{SEC} dataset following the publicly available recipe\footnote{\url{https://github.com/qiuqiangkong/audioset_tagging_cnn}}. 
The base model consisted of the CNN14 network trained on the AudioSet dataset~\cite{gemmeke2017audio}, where we replaced the output layer to
classify the 61 \gls{SE} classes of our experiments.
We retrained the model for 10,000 iterations using the Adam optimizer~\cite{kingma2015adam}, with a learning rate of $10^{-4}$ and a batch size of 32.

\subsection{Evaluation metrics}
We measured the extraction performance for active \gls{SE} classes in terms of \gls{SI-SDR} computed with the BSSEval toolkit~\cite{vincent2006performance} and reported the \gls{SDRi} compared to the mixtures.
\noteII{The \gls{SDRi} values were obtained by averaging the values for each \gls{SE} in the mixtures of the test set for the single-target extraction experiments. For the multi-target experiments, we extracted one combination of target \glspl{SE} per mixture for a number of targets $J$ of 1, 2, and 3.}

For inactive \gls{SE} classes, we measured the attenuation relative to the mixture, $\mathcal{A}^{\text{mix}}$, shown in Eq.~(\ref{eq:att_mix}) and relative to the minimum source, $\mathcal{A}^{\text{src}}$, which is computed by replacing $\y$ with the signal of the source having minimum power, $\x^{min}$, in Eq.~(\ref{eq:att_mix}). \noteII{The results show the average attenuation over inactive test samples (one randomly selected inactive \gls{SE} class per mixture).}  
Additionally, we evaluated the inactive class detection derived from the mixture attenuation ratio as explained in section \ref{sec:inactive} by plotting the \gls{ROC} curve (Recall vs. False positive rate) and measuring the \gls{AUC}.

Finally, for the evaluation of real mixtures, we used the \gls{SEC} model to predict the class labels after extraction and used the \gls{mAP} to measure the classification performance as it is widely used for \gls{AT}~\cite{kong2020panns}. \noteII{\gls{mAP} values were obtained by first computing the average precision (area under the recall-precision curve) per \gls{SE} class and then taking the average over the classes.} 
The \gls{mAP} value serves as a proxy to measure the extraction performance when the target source signals are unavailable.

\subsection{Results of single-target experiments} \label{sec:exp_single}

\begin{table}[t]
  \caption{\gls{SDRi} [dB] for single-target class experiment (i.e. J=1). 
  \gls{SDR} of the mixtures was -3.6 dB for both test sets. Bold font indicates the best performance, excluding oracle separation.}
  \label{tab:result_single}
  \centering
  \begin{tabular}{@{}c@{ }l@{ }c c c@{}}
\toprule
&Model & No. classes & \multicolumn{2}{c@{}}{No. classes (Test)} \\
&       & (Training)   & 41      & 61 \\
\midrule
1&\it{\gls{USS} (oracle selection)}   & 41 & \it{11.1} & \it{9.9} \\
2& \it{\gls{USS} (oracle selection)}                       & 61 & \it{10.6} & \it{9.8} \\
\midrule
3&\gls{USS} (\gls{SEC}-based selection) & 41 & 7.7 & 6.4 \\
4& \gls{USS} (\gls{SEC}-based selection)                    & 61 & 7.3 & 6.2 \\
\midrule
\midrule
5&Enrl    & 41        & 9.8  & 7.7 \\
6& Enrl       & 61        & 10.1  & 7.9  \\
\midrule
7&Class    & 41        & 10.9  & -  \\
8&  Class      & 61        & 10.3  & 8.3  \\
\midrule
\midrule
9&SoundBeam (Enrl)   & 41        & 9.9  & 7.9 \\
10& SoundBeam (Enrl) & 61        & 9.9  & 7.6  \\
\midrule
11&SoundBeam (Class)  & 41        & \bf{11.9}  & -  \\
12& SoundBeam (Class)  & 61        & 11.1   & 9.2  \\
\midrule
\midrule
13&SoundBeam-adapt (Class), $K=10$ &41+20 & \bf{11.9} & \bf{9.5} \\
\bottomrule
  \end{tabular}
\end{table}
 
 We first compare the different \gls{TSE} frameworks and a \gls{USS} baseline on the single-target datasets. Table \ref{tab:result_single} shows the \gls{SDRi} for the systems trained and tested on the single-target datasets with 41 and 61 \gls{SE} classes. 
 
 The first two systems (rows 1-4) consist of \gls{USS}-based approaches, which first separate the mixtures into three estimated source signals and then choose the target \gls{SE} class among the separated signals. The ``oracle selection'' (rows 1-2) chooses the separated signal with the highest SDR value. The ``\gls{SEC}-based selection'' (rows 3-4) chooses the signal with the highest posterior probability of belonging to the target class according to the \gls{SEC} model. The ``oracle selection'' represents an upper bound for \gls{USS}-based approaches, while the ``\gls{SEC}-based selection'' provides a more realistic performance level for cascade \gls{USS} and \gls{SEC} systems. We see that separating the sound mixtures with a high \gls{SDRi} of about 10 dB is possible. However, identifying the target \gls{SE} class from the separated output with an \gls{SEC} is more challenging, which results in a large performance drop of more than 3 dB. 
We could improve the cascade system by retraining the \gls{SEC} on the extracted signal, but it would not outperform the oracle selection.  

``Enrl'' (rows 5-6) and ``Class'' (rows 7-8) are the class label and enrollment-based \gls{TSE} systems, respectively.  ``SoundBeam (Enrl)'' (rows 9-10)  refers to the SoundBeam model using the enrollment encoder at test time. ``SoundBeam (Class)'' (rows 11-12) is the same model using the class label encoder at test time, and ``SoundBeam-adapt (Class)'' (row 13) is the SoundBeam (Class) model adapted to the new \gls{SE} classes.  
 All \gls{TSE} models outperform the ``\gls{USS} (\gls{SEC}-based selection)'' system, which demonstrates the potential of models optimized for \gls{TSE}. In the remaining experiments, we focus on \gls{TSE} and omit further comparison with \gls{USS}-based systems.
 
 The class label models outperform enrollment-based models because they can directly optimize the target \gls{SE} embedding vectors for each \gls{SE} class. ``SoundBeam (Class)'' (rows 11-12) performs the best and even outperforms the ``\gls{USS} (oracle selection)'' for the test set with 41 \gls{SE} classes. This result may be due to the multi-task training effect that seems to help to learn better class label embeddings.

Comparing the results with the 41- and 61-class test sets in Table \ref{tab:result_single}, we observe that the \gls{SDRi} is consistently worse with 61 classes than with 41 classes. This is even the case for the ``\gls{USS} (oracle selection)'' baseline (rows 1-2), 
indicating that it is more difficult to separate the mixtures as well.
There are two possible reasons for this issue. First, the additional 20 \gls{SE} classes may have sound characteristics that are more difficult to model. \noteII{Second, the training data contains fewer samples from these \gls{SE} classes since they are less represented in the original \gls{FSD} datasets} (i.e., on average 220 samples per \gls{SE} class for the first 41 \gls{SE} classes, but only 47 for the remaining 20 classes).

 \subsection{Results of few-shot adaptation on new SE classes}
 The results of Table \ref{tab:result_single}  confirm the ability of enrollment-based approaches trained with the dataset of 41 \gls{SE} classes (rows 5 and 9) to extract new classes from the test set with 61 \gls{SE} classes.
 However, \gls{SDRi} is lower by between 0.6 dB and 1.3 dB than when using the class label approaches trained with 61 classes (rows 8 and 12).
 
 Row 13 of Table \ref{tab:result_single} shows the results obtained using SoundBeam with few-shot adaptation to the 20 new \gls{SE} classes as described in section \ref{sec:new_SE}. The SoundBeam-adapt model was first trained with the dataset with 41 \gls{SE} classes and then adapted to the 20 new \gls{SE} classes using adaptation data generated from $K=10$ enrollments from each new \gls{SE} class. To simplify the experiments, we performed adaptation simultaneously for 20 new \gls{SE} classes. Since we only update the embedding matrix, this is equivalent to performing adaptation separately for each new \gls{SE} class.
 
With the few-shot adaptation scheme, we can directly optimize the embedding vectors for the new \gls{SE} classes while preserving performance on the seen \gls{SE} classes. Interestingly, the SoundBeam-adapt model performs slightly better than SoundBeam trained on 61 classes from scratch. These results confirm the effectiveness of the proposed adaptation scheme even if it uses fewer samples to learn the new classes (i.e., on average, 47 samples when training from scratch compared with 10 for the few-shot adaptation).

 \begin{figure*}[t]
  \centering
  \centerline{\includegraphics[width=0.95\linewidth]{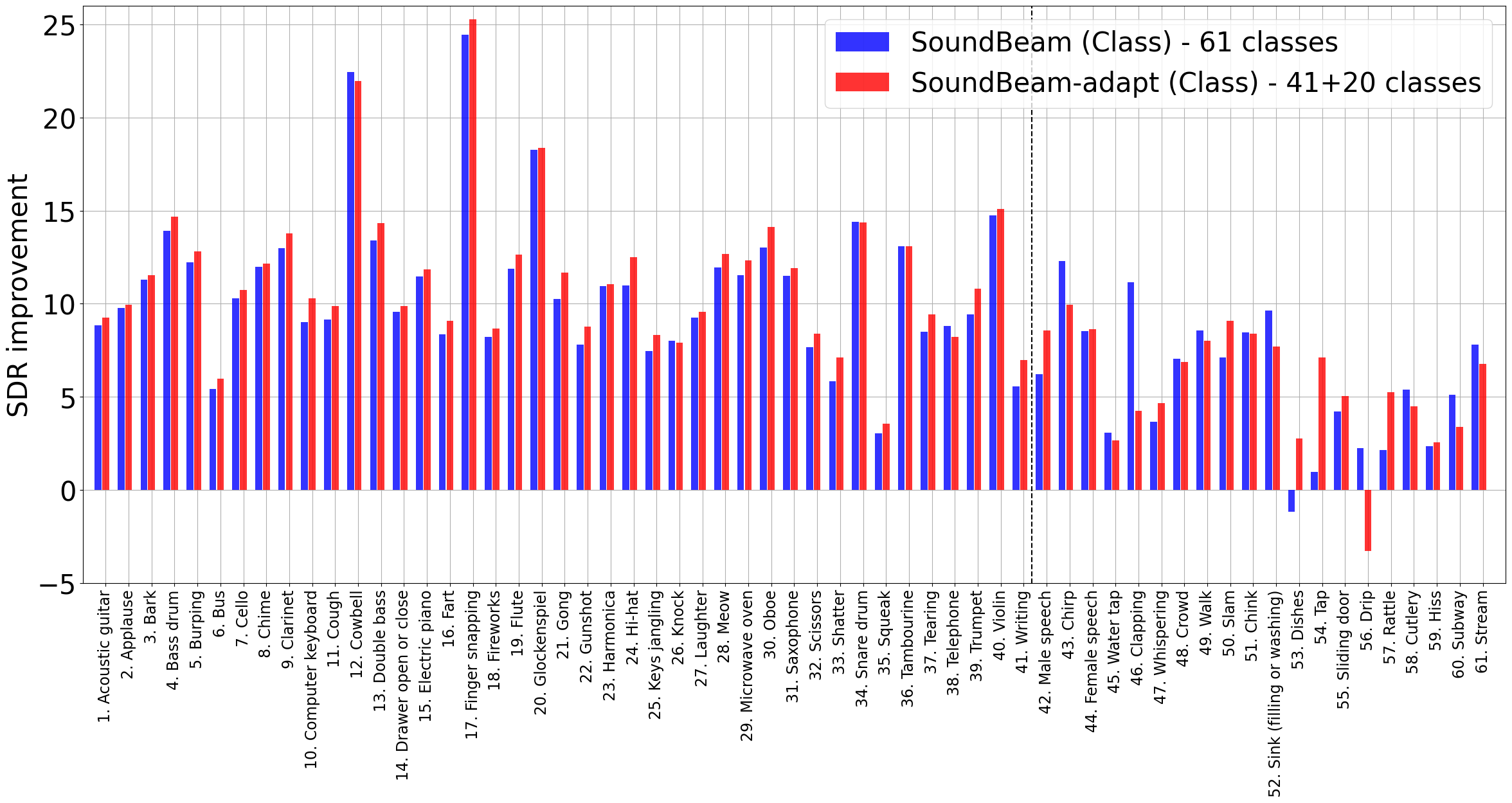}}
%
\caption{\gls{SDRi} [dB] for each \gls{SE} class for SoundBeam models without and with few-shot adaptation, i.e. model trained with the single-target datasets with 61 classes (SoundBeam (Class)) and model trained with 41 classes then adapted to 20 new \gls{SE} classes using the adaptation data (SoundBeam-adapt (Class).}
\label{fig:res_per_class}
%
\end{figure*}

Figure \ref{fig:res_per_class} plots the \gls{SDRi}  for the different target \gls{SE} classes for the SoundBeam model with adaptation (row 13 of Table \ref{tab:result_single}) and without adaptation (row 12 of Table \ref{tab:result_single}). The first 41 classes are those in the 41 \gls{SE} class dataset. ``SoundBeam(Class) 61 classes'' was trained on the dataset with 61 \gls{SE} classes from scratch.
For the first 41 classes, the adapted model behaves identically to the model trained on the 41-class dataset (row 11 of Table \ref{tab:result_single}) since adaptation does not modify the embedding matrix for the seen \gls{SE} classes.

We observe that, overall, ``SoundBeam-Adapt'' provides comparative performance to the ``SoundBeam(Class) 61 classes'' model  for the last 20 \gls{SE} classes. However, it performs slightly better for most of the first 41 \gls{SE} classes, which translates to better average performance as seen in Table~\ref{tab:result_single}. \noteIII{Note, this behavior may change depending on the data used and especially the type of \gls{SE} classes and the amount of training samples from each class.} 

Figure \ref{fig:res_per_class} shows that SoundBeam can successfully extract the target sounds with \gls{SDRi} of more than 5 dB for most target \gls{SE} classes.
We also confirm that extraction tends to be more challenging for the 20 new \gls{SE} classes (especially for classes 53, 54, 56, 57, and 59), even when the training data included these classes.

\begin{figure}[t]
  \centering
  \centerline{\includegraphics[width=0.9\linewidth]{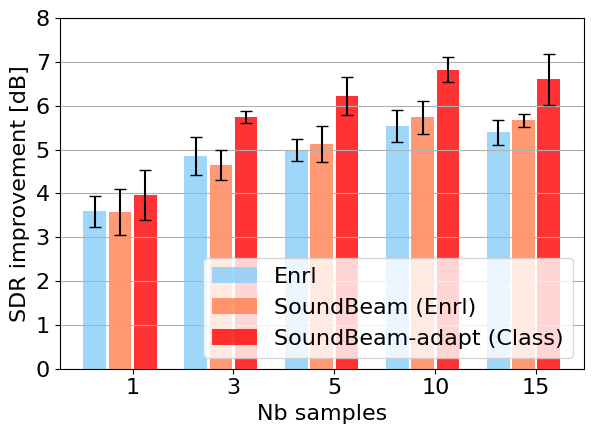}}
%
\caption{\gls{SDRi} as a function of the number of enrollments, $K$, for the Enrollment model, SoundBeam (Enrl) model trained with 41 \gls{SE} classes, and SoundBeam-adapt (Class) model. The figure shows the mean and variance over six trials of randomly selected enrollments. 
}
\label{fig:res_adaptation}
%
\end{figure}

Finally, we investigated the impact of the number of enrollment audio samples on the adaptation performance. Figure \ref{fig:res_adaptation} shows the \gls{SDR} improvement for the extraction of the new \gls{SE} classes as a function of the number of enrollment samples, $K$, for the enrollment-based \gls{TSE} model (``Enrl''), SoundBeam (``SoundBeam (Enrl)''), and SoundBeam with adaptation (``SoundBeam-adapt (Class)''). 
``Enrl'' and ``SoundBeam (Enrl)'' used the averaged embedding vectors of Eq. (\ref{eq:avg_emb}) to perform the extraction. Furthermore, SoundBeam with adaptation fine-tuned the averaged embeddings with the adaptation data.
We performed experiments over six trials of randomly selected enrollments and here report results on only the new \gls{SE} classes.

Figure~\ref{fig:res_adaptation} demonstrates that the proposed adaptation can improve performance by about 1 dB with only a few enrollments (i.e., $K \ge 3$). 
As a comparison, we also experimented with training the new embedding vectors from randomly initialized values, but this led to negligible \gls{SDRi}.  

The experiments with new \gls{SE} classes reveal that SoundBeam, with the adaptation scheme, can optimize the embedding vectors for the new \gls{SE} classes, which leads to improved performance. This result demonstrates the potential of SoundBeam for continuous learning of new \gls{SE} classes.

\subsection{Results for inactive SE classes}
\begin{table}[t]
  \caption{Results for inactive class extraction experiment for \gls{TSE} models trained with datasets having 61 \gls{SE} classes.} 
  \label{tab:result_inactive}
  \centering
  \begin{tabular}{@{}l@{ }c@{ }c@{ }c@{ }c@{ }c  c@{ }c@{ }c@{ }c@{}}
\toprule
Model & IS  & \multicolumn{4}{c}{41 test classes} &\multicolumn{4}{c}{61 test classes} \\
      &     &  $\mathcal{A}^{\text{mix}} \downarrow$ & $\mathcal{A}^{\text{src}} \downarrow$ & AUC $\uparrow$& SDRi $\uparrow$ &  $\mathcal{A}^{\text{mix}}\downarrow$ & $\mathcal{A}^{\text{src}}\downarrow$ & AUC $\uparrow$ &SDRi $\uparrow$\\
\midrule
 Enrl &  -  & -10.1 & -0.2 & 0.62 & 10.1 & -10.3 & -0.5 & 0.60 & 7.9\\
     &  \checkmark & -31.5 & -21.6 & 0.79 & 9.4 & -31.7 & -21.9 & 0.74 & 6.6   \\ 
\midrule
Class &  -  & -17.6 & -7.7 & 0.77 & 10.3  & -18.2 & -8.4 & 0.75  & 8.3 \\ 
     &  \checkmark & -40.2 & -30.3 & 0.89 & 9.6 & -42.7 & -32.9 & \bf{0.85}  & 7.2  \\ 
\midrule
SoundBeam &  -  & -11.1 & -1.1 & 0.65 & 9.9 & -11.1 & -1.3 & 0.62 & 7.6  \\ 
(Enrl) &  \checkmark & -30.2 & -20.3 & 0.79 & 9.4  & -30.3 & -20.5 & 0.75 & 6.8 \\ 
\midrule
SoundBeam &  -  & -16.0 & -6.1 & 0.74 & \bf{11.1} & -16.9 & -7.1 & 0.72 & \bf{9.2}  \\ 
(Class) &  \checkmark & \bf{-49.6} & \bf{-39.7} & \bf{0.90} &10.3  & \bf{-52.2} & \bf{-42.4} & 0.84  &  7.1 \\ 
\bottomrule
  \end{tabular}
\end{table}

\begin{figure}[t]
  \centering
  \centerline{\includegraphics[width=0.8\linewidth]{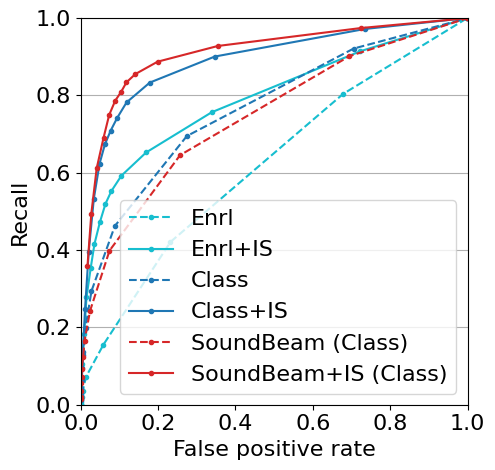}}
%
\caption{ROC curves for inactive event detection using different \gls{TSE} models. Results are shown for the test set with 41 classes.}
\label{fig:inactive detection}
%
\end{figure}

The next experiment investigates \gls{TSE} of inactive \gls{SE} classes as discussed in Section \ref{sec:inactive}. Table \ref{tab:result_inactive} compares the different \gls{TSE} models trained with and without \gls{IS} on the dataset with 61 \gls{SE} classes and tested on the evaluation data with 41 and 61 \gls{SE} classes.
In this experiment, for each mixture in the test sets, we randomly chose one \textit{inactive} \gls{SE} class and computed the attenuation and \gls{AUC} for that class. We report the averaged results in the table.
$\mathcal{A}^{\text{mix}}$ and $\mathcal{A}^{\text{src}}$ are the attenuation with respect to the mixture and the lowest source. In the table, $\downarrow$ indicates that the lower the attenuation, the better the model outputs zero signals for inactive classes.
The \gls{AUC} measures the performance of inactive \gls{SE} class detection using a simple classifier that considers a source inactive when $\mathcal{A}^{\text{mix}}$ is lower than a threshold. $\uparrow$ indicates that the higher the \gls{AUC}, the better the model can detect inactive classes. Figure \ref{fig:inactive detection} shows the \gls{ROC} curves that were used to compute the \gls{AUC}. Finally, we also report, in Table \ref{tab:result_inactive}, the \gls{SDRi} for the extraction of the \textit{active} \gls{SE} classes in the mixture. 

The results of Table \ref{tab:result_inactive} confirm that models trained without \gls{IS} cannot detect inactive \gls{SE} classes well and thus cannot output zero signals. In particular, the enrollment-based models perform extraction based on the similarities of the enrollments and sounds in the mixture. Consequently, if not explicitly taught to output zero signals for inactive sources, they tend to pick up the closest sounds in the mixture leading to a $\mathcal{A}^{\text{src}}$ value of around 0 dB. 
When using \gls{IS}, which adds 10 \% of inactive \gls{SE} classes to the training data as discussed in Section \ref{sec:inactive}, all models detect inactive \gls{SE} classes better with $\mathcal{A}^{\text{src}}$ values below -20 dB and \gls{AUC} values above 0.75 in all cases. 

Overall, the class label and SoundBeam models with \gls{IS} can identify inactive \gls{SE} classes well. Note that detecting inactive classes comes at the expense of lower extraction performance for the active \gls{SE} classes, especially for the test set with 61 \gls{SE} classes. 
As discussed previously, the 61 \gls{SE} class test set contains classes that are more challenging to identify, and thus the model trained with \gls{IS} tends to more often consider an active class as inactive. In future works, we will consider approaches to mitigate this issue by training better models for the 61 \gls{SE} classes using, for example, training data augmentation for the challenging \gls{SE} classes, as well as better tuning the amount of \gls{IS} during training.
\noteIII{Besides, some works dealing with target speech extraction\cite{Zhang2021,Delcroix2022} have recently proposed to address the inactive speaker case by using a speaker verification module to confirm that the extracted signal was really from the target speaker and otherwise considering the target as inactive. In future works, we will explore similar verification-based ideas for the \gls{TSE} problem.}

\subsection{Results of multi-target experiments}

\begin{table}[t]
  \caption{\gls{SDR} [dB] for multi-target extraction. Numbers in parentheses show the \gls{SDRi} compared to the mixture.}
  \label{tab:result_multi}
  \centering
  \begin{tabular}{l c ccc }
\toprule
Model & No. of  & \multicolumn{3}{c}{No. of sources in mixture $M$} \\
& targets $J$ & 3 & 4 & 5 \\
\midrule
Mixture &  1 & -3.6 & -5.6 & -6.8   \\ 
         &  2 & 4.0 & 0.2 & -2.0   \\ 
         &  3 & 21.0 & 6.0 & 2.3   \\ 
\midrule
Enrl &  1 & 3.2 (6.8) & 0.6 (6.2) & -1.0 (5.8)   \\ 
         &  2 & 7.9 (3.9) & 3.6 (3.5) & 1.4 (3.4)   \\ 
         &  3 & 21.3 (0.3) & 7.6 (1.5) & 3.9 (1.7)   \\ 
\midrule
Class &  1 & 4.4 (8.0) & 2.4 (7.9) & 0.8 (7.6)  \\ 
         &  2 & 8.3 (4.3) & \bf{5.2 (5.0)} & 2.9 (4.9)   \\ 
         &  3 & 17.2 (-3.7) & 8.6 (2.5) & 5.4 (3.1)   \\ 
\midrule
SoundBeam (Enrl) &  1 & 2.9 (6.5) & 0.3 (5.8) & -1.3 (5.5)   \\ 
         &  2 & 7.8 (3.8) & 3.5 (3.3) & 1.2 (3.2)   \\ 
         &  3 & \bf{21.6 (0.6)} & 7.5 (1.4) & 3.8 (1.6)  \\ 
\midrule
SoundBeam (Class) &  1 & \bf{4.9 (8.5)} & \bf{2.8 (8.3)} & \bf{1.1 (7.9)}   \\ 
         &  2 & \bf{8.8 (4.8)} & \bf{5.2 (5.0)} & \bf{3.0 (5.0)}   \\ 
         &  3 & 18.2 (-2.8) & \bf{8.9 (2.9)} & \bf{5.5 (3.2)}   \\
\bottomrule
  \end{tabular}
\end{table}

We tested the different models for the simultaneous extraction of multiple target \gls{SE} classes as discussed in Section \ref{sec:multiclass}.
It is also possible to extract multiple-target \gls{SE} classes using an iterative scheme, which extracts each class at a time. We showed in an earlier work~\cite{Ochiai2020} that both schemes achieved comparable performance, but the iterative scheme is more computationally expensive. Therefore, we omit here the results with the iterative scheme.

Table \ref{tab:result_multi} compares the \gls{SDR} values for various numbers of sources in the mixtures ($M$ from 3 to 5) and targets ($J$ from 1 to 3) for the multi-target dataset described in Section \ref{ssec:multi-dataset}. The top three rows indicate the SDR of the mixtures, which, as expected, increases with the number of target \gls{SE} classes. In the extreme case of extracting sounds from three \gls{SE} classes, $J=3$, in mixtures of three \gls{SE} sounds, $M=3$, the mixture and the reference, $\xS$, differ only by the presence of background noise, and thus the \gls{SDR} of the mixture attains 21 dB. It is thus naturally more challenging to improve \gls{SDR} for multi-target extraction. 

The results of Table \ref{tab:result_multi} confirm that simultaneous multi-target extraction is possible. Here again, SoundBeam outperforms the enrollment- or class label-based models.

Interestingly, we also observe that the extraction performance is not greatly affected by the number of sources in the mixture, i.e., we achieved similar \gls{SDRi} values for the mixtures of 3 to 5 sources. This result confirms that \gls{TSE} can operate independently of the number of sources in the mixture. Note that in our previous work~\cite{Ochiai2020}, we also confirmed that  \gls{TSE} was possible when the number of sources in the mixtures was not encountered during training.

\subsection{Investigations with real mixtures}
Finally, we perform preliminary experiments to explore the challenge of \gls{TSE} when using real mixtures as discussed in Section \ref{sec:realmixtures}.
\subsubsection{Measuring performance with SEC}
\begin{table}[t]
  \caption{\gls{SDRi} and \gls{mAP} values for single-target experiment on simulated data. Performing \gls{SEC} directly on the mixtures achieved  \gls{mAP}(tgt) and \gls{mAP}(all) 
  of \noteII{0.13 and 0.39}, respectively.}
  \label{tab:result_classification}
  \centering
  \begin{tabular}{l c c c c}
\toprule
Model & \multicolumn{2}{c}{41 test classes} & \multicolumn{2}{c}{61 test classes} \\
& SDRi & mAP & SDRi & mAP \\
\midrule
Enrl & 10.1 & \noteII{0.46} & 7.9 & \noteII{0.36}   \\ 
Class & 10.3 & \noteII{0.53} & 8.3 & \noteII{0.45}   \\ 
SoundBeam (Enrl) & 9.9 & \noteII{0.46} & 7.6 & \noteII{0.35}   \\ 
SoundBeam (Class) & 11.1 & \noteII{0.55} & 9.2 & \noteII{0.47}  \\ 
\bottomrule
  \end{tabular}
\end{table}

\begin{figure}[t]
  \centering
  \centerline{\includegraphics[width=0.8\linewidth]{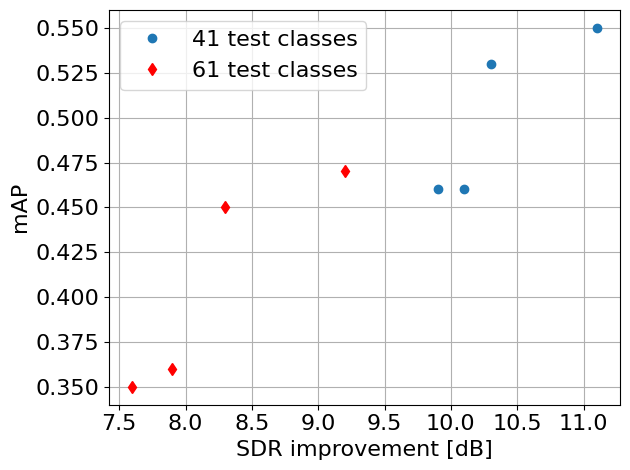}}
%
\caption{\gls{mAP} as a function of the \gls{SDRi} for various models.}
\label{fig:sdr_precision}
%
\end{figure}

For real recordings, we do not have access to the isolated target \gls{SE} signals, $\x^s$. It is thus not possible to compute SDR values. Therefore, to evaluate the performance on real recordings, we propose measuring the classification performance of \gls{SEC} applied to the extracted signals.
A similar approach was recently proposed for evaluating speech separation performance~\cite{maciejewski21_interspeech}. 

First, we evaluate how well we can measure \gls{TSE} performance with \gls{SEC} by comparing \gls{SDRi} and \gls{mAP} on \textit{simulated mixtures}.
Table~\ref{tab:result_classification} shows the \gls{SDRi} and \gls{mAP}  obtained by classifying the output of \gls{TSE} for the simulated single-target dataset. We should compare the \gls{mAP} values with those obtained by classifying the mixture directly. When scoring the mixture against the target \gls{SE} class (i.e., considering one target class as the reference), we obtain a \gls{mAP}(tgt) score of 0.13.
When scoring the mixture against all active \gls{SE} classes in the mixture (i.e., considering all active classes as the references), we obtain a \gls{mAP}(all) score of 0.39.
Here, we only report \gls{mAP}(tgt) when applying \gls{SEC}  after extraction. 

We observe that the \gls{mAP} values increase greatly after \gls{TSE}, achieving values over 0.45 for most models. 
Figure \ref{fig:sdr_precision} plots the \gls{mAP} values as a function of the \gls{SDRi} for the different systems of Table \ref{tab:result_classification}. Although the number of systems is limited, the results suggest a relationship between \gls{SDRi} and \gls{mAP} values. This result justifies our use of \gls{mAP} to evaluate extraction performance on real mixtures when it is impossible to compute the \gls{SDR} values.

\subsubsection{Extraction results with real mixtures}

\begin{table}[t]
  \caption{\gls{mAP} values for experiment with real mixtures. Performing \gls{SEC} directly on the mixtures achieved  \gls{mAP}(tgt) and \gls{mAP}(all) values of \noteII{0.25 and 0.55},
  respectively.}
  \label{tab:result_real}
  \centering
  \begin{tabular}{l c}
\toprule
Model & \gls{mAP} \\
\midrule
SoundBeam (Class) & \noteII{0.47} \\
+ Weakly supervised retraining & \noteII{0.49} \\
\bottomrule
  \end{tabular}
\end{table}

Table \ref{tab:result_real} shows the \gls{mAP} values on the real mixtures for the SoundBeam model trained with the single-target dataset with 61 \gls{SE} classes.
The \gls{mAP} \noteII{increases from 0.25 to 0.47}, which indicates that \gls{SEC} performance improved after extraction, suggesting that SoundBeam could extract the target. 

There is a mismatch between the recording conditions of the real mixtures and the simulated training data.
We experimented with retraining the model on training data consisting of real mixtures, using the weakly supervised \gls{SEC} loss described in Section \ref{sec:realmixtures}. The last row of Table \ref{tab:result_real} shows the results after retraining, which improved \gls{mAP} by \noteII{2 points}. \noteII{This difference is significant according to a Student's paired t-test~\cite{Smucker_2007} for a \textit{p}-value of 0.059.}

We should emphasize that these results are only suggestive and should be considered with precaution because good \gls{SEC} does not always mean good extraction. For example, if only a portion of the sound of the target \gls{SE} class is extracted, the \gls{mAP} can be high, although the extraction would be imperfect. 

Although they provide a likely imperfect evaluation, the results of Table \ref{tab:result_real} combined with informal listening\footnote{\label{fn:demo}Sound samples can be found on our demo webpage \url{www.kecl.ntt.co.jp/icl/signal/member/marcd/SoundBeamDemo}} indicate that SoundBeam can perform \gls{TSE} on real mixtures. However, they also imply that future work is required to improve extraction performance and measure performance more accurately.

\section{Conclusion}
\label{sec:conclusion}
In this paper, we introduced a \gls{TSE} framework, SoundBeam, and performed extensive experiments comparing it to enrollment-based and class label-based  schemes. We showed that the SoundBeam model combined the strengths of both enrollment- and class label-based schemes, which translated to better overall performance in various conditions, including inactive classes, new classes, and multi-target extraction. Furthermore, it offers the possibility of learning how to extract \gls{SE} classes with few-shot adaptation.
We also discussed the applicability of SoundBeam to processing recordings of real sound mixtures.
Sound samples of the proposed SoundBeam are available on our demo webpage\footref{fn:demo}

These experiments show the potential of \gls{TSE} to tackle practical applications. However, there are many remaining issues. First, we focused on offline processing using computationally intensive network configurations, but many applications of \gls{TSE} would require online processing with limited computational resources. The ideas presented could also be applied to a causal implementation of the models~\cite{luo2019conv} or more efficient models~\cite{tzinis2020sudo}, but more investigations would be required.
Second, although we demonstrated promising initial results on real recordings, further research is still needed to improve extraction performance. For example, future works could include designing simulated training data that approximate better mixing conditions of real mixtures, exploiting larger datasets for semi-supervised retraining, or combining mixtures-of-mixtures and \gls{SEC}-based training/adaptation strategies.
Finally, we would like to extend our investigations to an even larger number of \gls{SE} classes using larger datasets such as \noteII{AudioSet}~\cite{gemmeke2017audio}.


\ifCLASSOPTIONcaptionsoff
  \newpage
\fi



%
\bibliographystyle{IEEEtran}
\bibliography{mybib}

\begin{thebibliography}{10}
\providecommand{\url}[1]{#1}
\csname url@samestyle\endcsname
\providecommand{\newblock}{\relax}
\providecommand{\bibinfo}[2]{#2}
\providecommand{\BIBentrySTDinterwordspacing}{\spaceskip=0pt\relax}
\providecommand{\BIBentryALTinterwordstretchfactor}{4}
\providecommand{\BIBentryALTinterwordspacing}{\spaceskip=\fontdimen2\font plus
\BIBentryALTinterwordstretchfactor\fontdimen3\font minus
  \fontdimen4\font\relax}
\providecommand{\BIBforeignlanguage}[2]{{%
\expandafter\ifx\csname l@#1\endcsname\relax
\typeout{** WARNING: IEEEtran.bst: No hyphenation pattern has been}%
\typeout{** loaded for the language `#1'. Using the pattern for}%
\typeout{** the default language instead.}%
\else
\language=\csname l@#1\endcsname
\fi
#2}}
\providecommand{\BIBdecl}{\relax}
\BIBdecl

\bibitem{cherry_1953}
E.~C. Cherry, ``Some experiments on the recognition of speech, with one and
  with two ears,'' \emph{JASA}, vol.~25, no.~5, pp. 975--979, 1953.

\bibitem{Ochiai2020}
T.~Ochiai, M.~Delcroix, Y.~Koizumi, H.~Ito, K.~Kinoshita, and S.~Araki,
  ``Listen to what you want: Neural network-based universal sound selector,''
  in \emph{Proc. of Interspeech}, 2020, pp. 1441--1445.

\bibitem{cakir2015polyphonic}
E.~Cakir, T.~Heittola, H.~Huttunen, and T.~Virtanen, ``Polyphonic sound event
  detection using multi label deep neural networks,'' in \emph{Proc. of IJCNN},
  2015, pp. 1--7.

\bibitem{mesaros2021sound}
A.~Mesaros, T.~Heittola, T.~Virtanen, and M.~D. Plumbley, ``Sound event
  detection: A tutorial,'' \emph{IEEE Signal Processing Magazine}, vol.~38,
  no.~5, pp. 67--83, 2021.

\bibitem{hershey2017cnn}
S.~Hershey, S.~Chaudhuri, D.~P. Ellis, J.~F. Gemmeke, A.~Jansen, R.~C. Moore,
  M.~Plakal, D.~Platt, R.~A. Saurous, B.~Seybold \emph{et~al.}, ``{CNN}
  architectures for large-scale audio classification,'' in \emph{Proc. of
  ICASSP}, 2017, pp. 131--135.

\bibitem{kong2020panns}
Q.~Kong, Y.~Cao, T.~Iqbal, Y.~Wang, W.~Wang, and M.~D. Plumbley, ``Panns:
  Large-scale pretrained audio neural networks for audio pattern recognition,''
  \emph{IEEE/ACM Trans. ASLP}, vol.~28, pp. 2880--2894, 2020.

\bibitem{clavel2005events}
C.~Clavel, T.~Ehrette, and G.~Richard, ``Events detection for an audio-based
  surveillance system,'' in \emph{Proc. of ICME}, 2005, pp. 1306--1309.

\bibitem{atrey2006audio}
P.~Atrey, N.~Maddage, and M.~Kankanhalli, ``Audio based event detection for
  multimedia surveillance,'' in \emph{Proc. of ICASSP}, 2006, pp. 813--816.

\bibitem{mesaros2017dcase}
A.~Mesaros, T.~Heittola, A.~Diment, B.~Elizalde, A.~Shah, E.~Vincent, B.~Raj,
  and T.~Virtanen, ``{DCASE} 2017 {Challenge} setup: {Tasks}, datasets and
  baseline system,'' in \emph{Proc. of DCASE}, 2017.

\bibitem{kavalerov2019universal}
I.~Kavalerov, S.~Wisdom, H.~Erdogan, B.~Patton, K.~Wilson, J.~Le~Roux, and
  J.~R. Hershey, ``Universal sound separation,'' in \emph{Proc. of WASPAA},
  2019, pp. 175--179.

\bibitem{kong2020source}
Q.~Kong, Y.~Wang, X.~Song, Y.~Cao, W.~Wang, and M.~D. Plumbley, ``Source
  separation with weakly labelled data: An approach to computational auditory
  scene analysis,'' in \emph{Proc. of ICASSP}, 2020, pp. 101--105.

\bibitem{gfeller2020oneshot}
B.~Gfeller, D.~Roblek, and M.~Tagliasacchi, ``One-shot conditional audio
  filtering of arbitrary sounds,'' in \emph{Proc. of ICASSP}, 2021, pp.
  501--505.

\bibitem{delcroix2021few}
M.~Delcroix, J.~B. V{\'a}zquez, T.~Ochiai, K.~Kinoshita, and S.~Araki,
  ``Few-shot learning of new sound classes for target sound extraction,''
  \emph{Proc. of Interspeech}, 2021.

\bibitem{okamoto2021environmental}
Y.~Okamoto, S.~Horiguchi, M.~Yamamoto, K.~Imoto, and Y.~Kawaguchi,
  ``Environmental sound extraction using onomatopoeia,'' \emph{arXiv preprint
  arXiv:2112.00209}, 2021.

\bibitem{chen2022zero}
K.~Chen, X.~Du, B.~Zhu, Z.~Ma, T.~Berg-Kirkpatrick, and S.~Dubnov, ``Zero-shot
  audio source separation through query-based learning from weakly-labeled
  data,'' in \emph{Proc. of the AAAI}, vol.~36, no.~4, 2022, pp. 4441--4449.

\bibitem{LeeCL19}
J.~H. Lee, H.~Choi, and K.~Lee, ``Audio query-based music source separation,''
  in \emph{Proc. of ISMIR}, 2019, pp. 878--885.

\bibitem{slizovskaia2019end}
O.~Slizovskaia, L.~Kim, G.~Haro, and E.~Gomez, ``End-to-end sound source
  separation conditioned on instrument labels,'' in \emph{Proc. of ICASSP},
  2019, pp. 306--310.

\bibitem{zmolikova2019speakerbeam}
K.~Zmolikova, M.~Delcroix, K.~Kinoshita, T.~Ochiai, T.~Nakatani, L.~Burget, and
  J.~Cernocky, ``{SpeakerBeam}: Speaker aware neural network for target speaker
  extraction in speech mixtures,'' \emph{IEEE JSTSP}, vol.~13, no.~4, pp.
  800--814, 2019.

\bibitem{seetharaman2019class}
P.~Seetharaman, G.~Wichern, S.~Venkataramani, and J.~Le~Roux,
  ``Class-conditional embeddings for music source separation,'' in \emph{Proc.
  of ICASSP}, 2019, pp. 301--305.

\bibitem{delcroix2018single}
M.~Delcroix, K.~Zmolikova, K.~Kinoshita, A.~Ogawa, and T.~Nakatani, ``Single
  channel target speaker extraction and recognition with {SpeakerBeam},'' in
  \emph{Proc. of ICASSP}, 2018, pp. 5554--5558.

\bibitem{fonseca2020fsd50k}
E.~Fonseca, X.~Favory, J.~Pons, F.~Font, and X.~Serra, ``{FSD50k}: an open
  dataset of human-labeled sound events,'' \emph{arXiv preprint
  arXiv:2010.00475}, 2020.

\bibitem{kolbaek2017multitalker}
M.~Kolb{\ae}k, D.~Yu, Z.-H. Tan, and J.~Jensen, ``Multitalker speech separation
  with utterance-level permutation invariant training of deep recurrent neural
  networks,'' \emph{IEEE/ACM Trans. ASLP}, vol.~25, no.~10, pp. 1901--1913,
  2017.

\bibitem{hershey2016deep}
J.~R. Hershey, Z.~Chen, J.~Le~Roux, and S.~Watanabe, ``Deep clustering:
  Discriminative embeddings for segmentation and separation,'' in \emph{Proc.
  of ICASSP}, 2016, pp. 31--35.

\bibitem{luo2019conv}
Y.~Luo and N.~Mesgarani, ``{Conv-TasNet}: Surpassing ideal time--frequency
  magnitude masking for speech separation,'' \emph{IEEE/ACM Trans. ASLP},
  vol.~27, no.~8, pp. 1256--1266, 2019.

\bibitem{Cano_2019}
E.~Cano, D.~FitzGerald, A.~Liutkus, M.~D. Plumbley, and F.-R. Stöter,
  ``Musical source separation: An introduction,'' \emph{IEEE Signal Processing
  Magazine}, vol.~36, no.~1, pp. 31--40, 2019.

\bibitem{Uhlich_icassp17}
S.~Uhlich, M.~Porcu, F.~Giron, M.~Enenkl, T.~Kemp, N.~Takahashi, and
  Y.~Mitsufuji, ``Improving music source separation based on deep neural
  networks through data augmentation and network blending,'' in \emph{Proc. of
  ICASSP}, 2017, pp. 261--265.

\bibitem{luo2017_deepclustering}
Y.~Luo, Z.~Chen, J.~R. Hershey, J.~Le~Roux, and N.~Mesgarani, ``Deep clustering
  and conventional networks for music separation: Stronger together,'' in
  \emph{Proc. of ICASSP}, 2017, pp. 61--65.

\bibitem{rafii2018overview}
Z.~Rafii, A.~Liutkus, F.-R. St{\"o}ter, S.~I. Mimilakis, D.~FitzGerald, and
  B.~Pardo, ``An overview of lead and accompaniment separation in music,''
  \emph{IEEE/ACM Trans. ASLP}, vol.~26, no.~8, pp. 1307--1335, 2018.

\bibitem{kumar18c_interspeech}
R.~Kumar, Y.~Luo, and N.~Mesgarani, ``{Music Source Activity Detection and
  Separation Using Deep Attractor Network},'' in \emph{Proc. Interspeech},
  2018, pp. 347--351.

\bibitem{FUSS_wisdom2021s}
S.~Wisdom, H.~Erdogan, D.~P. Ellis, R.~Serizel, N.~Turpault, E.~Fonseca,
  J.~Salamon, P.~Seetharaman, and J.~R. Hershey, ``What’s all the {FUSS}
  about free universal sound separation data?'' in \emph{Proc. of ICASSP},
  2021, pp. 186--190.

\bibitem{tzinis2020sudo}
E.~Tzinis, Z.~Wang, and P.~Smaragdis, ``Sudo rm-rf: Efficient networks for
  universal audio source separation,'' in \emph{Proc. of MLSP}, 2020, pp. 1--6.

\bibitem{olvera2021foreground}
M.~Olvera, E.~Vincent, R.~Serizel, and G.~Gasso, ``Foreground-background
  ambient sound scene separation,'' in \emph{Proc. of EUSIPCO}, 2021, pp.
  281--285.

\bibitem{heittola2011sound}
T.~Heittola, A.~Mesaros, T.~Virtanen, and A.~Eronen, ``Sound event detection in
  multisource environments using source separation,'' in \emph{Machine
  Listening in Multisource Environments}, 2011.

\bibitem{gemmeke2013exemplar}
J.~F. Gemmeke, L.~Vuegen, P.~Karsmakers, B.~Vanrumste \emph{et~al.}, ``An
  exemplar-based {NMF} approach to audio event detection,'' in \emph{Proc. of
  WASPAA}, 2013, pp. 1--4.

\bibitem{CornellDCASE2020}
S.~Cornell, G.~Pepe, E.~Principi, M.~Pariente, M.~Olvera, L.~Gabrielli, and
  S.~Squartini, ``The {UNIVPM-INRIA} systems for the {DCASE} 2020 task 4,''
  \emph{Proc. of DCASE}, 2020.

\bibitem{turpault2020improving}
N.~Turpault, S.~Wisdom, H.~Erdogan, J.~Hershey, R.~Serizel, E.~Fonseca,
  P.~Seetharaman, and J.~Salamon, ``Improving sound event detection in domestic
  environments using sound separation,'' \emph{Proc. of DCASE}, 2020.

\bibitem{huang2020guided}
Y.~Huang, L.~Lin, S.~Ma, X.~Wang, H.~Liu, Y.~Qian, M.~Liu, and K.~Ouch,
  ``Guided multi-branch learning systems for sound event detection with sound
  separation,'' \emph{Proc. of DCASE}, 2020.

\bibitem{Wang2019voicefilter}
Q.~Wang, H.~Muckenhirn, K.~Wilson, P.~Sridhar, Z.~Wu, J.~R. Hershey, R.~A.
  Saurous, R.~J. Weiss, Y.~Jia, and I.~L. Moreno, ``{VoiceFilter: Targeted
  Voice Separation by Speaker-Conditioned Spectrogram Masking},'' in
  \emph{Proc. of Interspeech}, 2019, pp. 2728--2732.

\bibitem{zhao2018sound}
H.~Zhao, C.~Gan, A.~Rouditchenko, C.~Vondrick, J.~McDermott, and A.~Torralba,
  ``The sound of pixels,'' in \emph{Proc. of ECCV}, 2018, pp. 570--586.

\bibitem{samuel2020meta}
D.~Samuel, A.~Ganeshan, and J.~Naradowsky, ``Meta-learning extractors for music
  source separation,'' in \emph{Proc. of ICASSP}, 2020, pp. 816--820.

\bibitem{zmolikova2017speaker}
K.~Zmolikova, M.~Delcroix, K.~Kinoshita, T.~Higuchi, A.~Ogawa, and T.~Nakatani,
  ``Speaker-aware neural network based beamformer for speaker extraction in
  speech mixtures,'' in \emph{Proc. of Interspeech}, 2017, pp. 2655--2659.

\bibitem{wisdom2020unsupervised}
S.~Wisdom, E.~Tzinis, H.~Erdogan, R.~J. Weiss, K.~Wilson, and J.~R. Hershey,
  ``Unsupervised sound separation using mixtures of mixtures,'' \emph{Proc. of
  NeurIPS}, 2020.

\bibitem{zhang2017weakly}
N.~Zhang, J.~Yan, and Y.~Zhou, ``Weakly supervised audio source separation via
  spectrum energy preserved wasserstein learning,'' \emph{Proc. of IJCAI-18},
  pp. 4574--4580, 7 2018.

\bibitem{stoller2018adversarial}
D.~Stoller, S.~Ewert, and S.~Dixon, ``Adversarial semi-supervised audio source
  separation applied to singing voice extraction,'' in \emph{Proc. of ICASSP},
  2018, pp. 2391--2395.

\bibitem{pishdadian2020finding}
F.~Pishdadian, G.~Wichern, and J.~Le~Roux, ``Finding strength in weakness:
  Learning to separate sounds with weak supervision,'' \emph{IEEE/ACM Trans.
  ASLP}, vol.~28, pp. 2386--2399, 2020.

\bibitem{gao2019co}
R.~Gao and K.~Grauman, ``Co-separating sounds of visual objects,'' in
  \emph{Proc. of ICCV}, 2019, pp. 3879--3888.

\bibitem{delcroix2020improving}
M.~Delcroix, T.~Ochiai, K.~Zmolikova, K.~Kinoshita, N.~Tawara, T.~Nakatani, and
  S.~Araki, ``Improving speaker discrimination of target speech extraction with
  time-domain {SpeakerBeam},'' in \emph{Proc. of ICASSP}, 2020, pp. 691--695.

\bibitem{xiao2019single}
X.~Xiao, Z.~Chen, T.~Yoshioka, H.~Erdogan, C.~Liu, D.~Dimitriadis, J.~Droppo,
  and Y.~Gong, ``Single-channel speech extraction using speaker inventory and
  attention network,'' in \emph{Proc. of ICASSP}, 2019, pp. 86--90.

\bibitem{vesely2016sequence}
K.~Vesely, S.~Watanabe, K.~Zmolikova, M.~Karafiat, L.~Burget, and J.~H.
  Cernocky, ``Sequence summarizing neural network for speaker adaptation,'' in
  \emph{Proc. of ICASSP}, 2016, pp. 5315--5319.

\bibitem{goodfellow2013empirical}
I.~J. Goodfellow, M.~Mirza, D.~Xiao, A.~Courville, and Y.~Bengio, ``An
  empirical investigation of catastrophic forgetting in gradient-based neural
  networks,'' \emph{Proc. of ICLR}, 2014.

\bibitem{borsdorf21_interspeech}
M.~Borsdorf, C.~Xu, H.~Li, and T.~Schultz, ``{Universal Speaker Extraction in
  the Presence and Absence of Target Speakers for Speech of One and Two
  Talkers},'' in \emph{Proc. Interspeech}, 2021, pp. 1469--1473.

\bibitem{fonseca2018general}
E.~Fonseca, M.~Plakal, F.~Font, D.~P. Ellis, X.~Favory, J.~Pons, and X.~Serra,
  ``General-purpose tagging of freesound audio with audioset labels: Task
  description, dataset, and baseline,'' in \emph{Proc. of DCASE}, 2018.

\bibitem{salamon2017scaper}
J.~Salamon, D.~MacConnell, M.~Cartwright, P.~Li, and J.~P. Bello, ``Scaper: A
  library for soundscape synthesis and augmentation,'' in \emph{Proc. of
  WASPAA}, 2017, pp. 344--348.

\bibitem{kinoshita2016summary}
K.~Kinoshita, M.~Delcroix, S.~Gannot, E.~A. Habets, R.~Haeb-Umbach,
  W.~Kellermann, V.~Leutnant, R.~Maas, T.~Nakatani, B.~Raj \emph{et~al.}, ``A
  summary of the {REVERB} challenge: state-of-the-art and remaining challenges
  in reverberant speech processing research,'' \emph{EURASIP Journal on
  Advances in Signal Processing}, vol.~7, 2016.

\bibitem{Pariente2020Asteroid}
M.~Pariente, S.~Cornell, J.~Cosentino, S.~Sivasankaran, E.~Tzinis,
  J.~Heitkaemper, M.~Olvera, F.-R. Stöter, M.~Hu, J.~M. Martin-Doñas,
  D.~Ditter, A.~Frank, A.~Deleforge, and E.~Vincent, ``{Asteroid: The
  PyTorch-Based Audio Source Separation Toolkit for Researchers},'' in
  \emph{Proc. of Interspeech}, 2020, pp. 2637--2641.

\bibitem{kingma2015adam}
D.~P. Kingma and J.~Ba, ``Adam: A method for stochastic optimization,'' in
  \emph{Proc. of ICLR}, 2015.

\bibitem{gemmeke2017audio}
J.~F. Gemmeke, D.~P. Ellis, D.~Freedman, A.~Jansen, W.~Lawrence, R.~C. Moore,
  M.~Plakal, and M.~Ritter, ``Audio set: An ontology and human-labeled dataset
  for audio events,'' in \emph{Proc. of ICASSP}, 2017, pp. 776--780.

\bibitem{vincent2006performance}
E.~Vincent, R.~Gribonval, and C.~Fevotte, ``Performance measurement in blind
  audio source separation,'' \emph{IEEE trans. ASLP}, vol.~14, no.~4, pp.
  1462--1469, 2006.

\bibitem{Zhang2021}
C.~Zhang, M.~Yu, C.~Weng, and D.~Yu, ``Towards robust speaker verification with
  target speaker enhancement,'' in \emph{Proc. of ICASSP'21}, 2021, pp.
  6693--6697.

\bibitem{Delcroix2022}
M.~Delcroix, K.~Kinoshita, T.~Ochiai, K.~Zmolikova, H.~Sato, and T.~Nakatani,
  ``Listen only to me! how well can target speech extraction handle false
  alarms?'' in \emph{Proc. of Interspeech'22}, 2022.

\bibitem{maciejewski21_interspeech}
M.~Maciejewski, S.~Watanabe, and S.~Khudanpur, ``{Speaker Verification-Based
  Evaluation of Single-Channel Speech Separation},'' in \emph{Proc.
  Interspeech}, 2021, pp. 3520--3524.

\bibitem{Smucker_2007}
\BIBentryALTinterwordspacing
M.~D. Smucker, J.~Allan, and B.~Carterette, ``A comparison of statistical
  significance tests for information retrieval evaluation,'' in
  \emph{Proceedings of the Sixteenth ACM Conference on Conference on
  Information and Knowledge Management}.\hskip 1em plus 0.5em minus 0.4em\relax
  New York, NY, USA: Association for Computing Machinery, 2007, p. 623–632.
  [Online]. Available: \url{https://doi.org/10.1145/1321440.1321528}
\BIBentrySTDinterwordspacing

\end{thebibliography}
\end{document}